\documentclass[preprintnumbers, prd,onecolumn, floatfix, superscriptaddress, nofootinbib]{revtex4-2}

\usepackage{setspace}
\usepackage{graphicx}
\usepackage{subcaption}
\usepackage{epsfig}
\usepackage{bm}
\usepackage{amssymb}
\usepackage{float}
\usepackage{amsmath}
\usepackage{dcolumn}
\usepackage[colorlinks]{hyperref}
\usepackage[usenames,dvipsnames]{color}
\usepackage{enumitem}

\usepackage[mathscr]{eucal}
\usepackage{orcidlink} % Simplest method for including ORCID iDs
\hypersetup{ breaklinks=true, pdfstartview={FitH}, colorlinks=true, linkcolor=blue, citecolor=red, filecolor=magenta, urlcolor=blue, anchorcolor=green, linktocpage=true }

\def\doi{http://doi.org}

\newcommand{\be}{\begin{equation}}
\newcommand{\ee}{\end{equation}}
\newcommand{\beano}{\begin{eqnarray*}}
\newcommand{\eeano}{\end{eqnarray*}}
\newcommand{\ba}{\begin{eqnarray}}
\newcommand{\ea}{\end{eqnarray}}

\begin{document}

\title{ Scale-Invariant Bounce Cosmology in Weyl $f(Q)$ Gravity with Quintom Signature}
\author{Rita Rani\orcidlink{0000-0001-8628-2368}}
\email{rita.ma19@nsut.ac.in}
\affiliation{Department of Mathematics, Netaji Subhas University of Technology, New Delhi-110 078, India}
\author{G. K. Goswami\orcidlink{0000-0002-2178-6925}}
\email{gk.goswami9@gmail.com}
\affiliation{Department of Mathematics, Netaji Subhas University of Technology, New Delhi-110 078, India}
\author{J. K. Singh\orcidlink{0009-0000-6037-6702}}
\email{jksingh@nsut.ac.in}
\affiliation{Department of Mathematics, Netaji Subhas University of Technology, New Delhi-110 078, India}
\author{Sushant G. Ghosh\orcidlink{0000-0002-0835-3690} }
\email{sghosh2@jmi.ac.in}
\affiliation{Centre for Theoretical Physics, Jamia Millia Islamia, New Delhi 110025, India}
\affiliation{Astrophysics and Cosmology Research Unit, School of Mathematics, Statistics and Computer Science, University of KwaZulu-Natal, Private Bag 54001, Durban 4000, South Africa}
\author{Sunil D Maharaj\orcidlink{0000-0002-1967-2849} }
\email{maharaj@ukzn.ac.za}
\affiliation{Astrophysics and Cosmology Research Unit, School of Mathematics, Statistics and Computer Science, University of KwaZulu-Natal, Private Bag 54001, Durban 4000, South Africa}

\begin{abstract}
\begin{singlespace}

We investigate a bouncing cosmological model within the Weyl-type $f(Q)$ gravity framework, employing a power-law form of the non-metricity scalar $Q$. The model successfully resolves the initial singularity problem by demonstrating a nonsingular bounce, where the universe transitions from a contracting phase $ \dot{a}(t)<0 $ to an expanding phase ($ \dot{a}(t)>0 $) at the bouncing point $t \approx 0.$ Key features include the violation of the null energy condition (NEC) near the bounce and the crossing of the phantom divide line ($\omega=-1$) by the equation of state (EoS) parameter, indicating quintom-like behavior. The model exhibits accelerated expansion post-bounce, suggesting an inflationary phase. Stability analysis via the adiabatic index reveals instability near the bouncing point, while energy conditions highlight the dominance of dark energy. Additionally, the study explores scalar fields, showing that quintessence-like kinetic energy becomes negative and phantom-like kinetic energy peaks positively near the bounce, aligning with dark energy dynamics. The Hubble parameter, deceleration parameter, and Hubble radius further validate the bouncing scenario, with the latter displaying symmetric behaviour around the bounce. These results underscore the viability of Weyl-type $f(Q)$ gravity as a framework for nonsingular bouncing cosmologies, offering insights into early universe dynamics and dark energy behaviour.

\end{singlespace}
\end{abstract}
 
\maketitle
PACS numbers: {98.80 cq, 04.50.Kd, 95.36.+x}\\

Keywords: Bouncing cosmology, Weyl-type $f(Q)$ gravity, quintom behavior,  Scalar field, Stability analysis

\section{Introduction} 

\qquad Modern cosmological observations have firmly established that our universe is currently undergoing accelerated expansion, as demonstrated by Type Ia supernova measurements \cite{SupernovaSearchTeam:1998fmf,perlmutter2003measuring} and precision cosmic microwave background data \cite{WMAP:2003elm, WMAP:2006bqn, WMAP:2003ogi, BICEP2:2014owc}. While Einstein's General Relativity (GR) revolutionised our understanding of gravity, its limitations in explaining cosmic acceleration and quantum-gravitational effects have motivated the development of alternative theories \cite{Clifton:2011jh}. Early modifications by Eddington, Weyl, and Kaluza-Klein laid the theoretical groundwork for contemporary extended gravity theories that either modify the energy-momentum content or generalize the geometric framework of the Einstein field equations \cite{Peebles:2002gy}. The need for such modifications becomes particularly apparent when addressing the dark universe paradigm, where dark energy (DE) drives the accelerated expansion, and dark matter (DM) explains galactic rotation curves \cite{Sahni:1999gb}. While GR successfully describes gravitational phenomena at solar system scales, its cosmological applications face significant challenges, including the cosmological constant problem and singularity theorems. This has led to the development of three principal geometric approaches to gravity: Riemannian (curvature-based), Weitzenböck (torsion-based), and symmetric teleparallel (non-metricity-based) formulations \cite{BeltranJimenez:2017tkd}, each offering unique perspectives on gravitational interactions.

The development of modified gravity theories has progressed significantly through several parallel approaches, each offering distinct geometric perspectives on gravitation. Curvature-based modifications like $f(R)$ gravity \cite{Starobinsky:2007hu, Capozziello:2008qc, Goswami:2022vfq} and its extensions to $f(R, G)$ \cite{Wu:2015maa, Naz:2023pfl, Rani:2024uah} and $f(R,T)$ \cite{ Harko:2011kv, Singh:2022eun, Singh:2024ckh, Singh:2024kez} theories have demonstrated success in addressing both early and late-time cosmic acceleration. Similarly, torsion-based formulations through $f(T)$ gravity \cite{Cai:2015emx, Capozziello:2011hj} and its generalizations \cite{Bahamonde:2015zma} have provided alternative teleparallel frameworks. Recently, $f(Q)$ gravity \cite{BeltranJimenez:2017tkd, Heisenberg:2023lru} has emerged as a particularly promising direction within symmetric teleparallel geometry, where gravity manifests through non-metricity rather than curvature. This approach naturally avoids ghost instabilities \cite{Lin:2021uqa} and maintains better agreement with cosmological observations \cite{Lazkoz:2019sjl, Frusciante:2021sio} compared to some alternatives. Current research explores its cosmological perturbations \cite{Mandal:2020lyq}, black hole solutions \cite{DAmbrosio:2021zpm}, thermodynamic properties \cite{Sokoliuk:2023pby}, and quantum aspects \cite{Bajardi:2023vcc}, while Weyl-type extensions \cite{Haghani:2012bt, Xu:2020yeg, Goswami:2023knh} further enrich the framework by incorporating scale-invariant vector fields. The $f(Q)$ paradigm continues to gain prominence as it offers novel solutions to persistent challenges in gravitational physics and cosmology. Some of the aspects of this theoretical framework that various researchers investigate include \cite{Lazkoz:2019sjl, Mandal:2020lyq, Frusciante:2021sio, Lin:2021uqa, DAmbrosio:2021zpm, Heisenberg:2023lru}.

While the standard cosmological model successfully describes many features of the early universe, several fundamental challenges remain unresolved, including the flatness and horizon problems, baryon asymmetry, and—most notably—the initial singularity. Although inflationary theory, pioneered by Alan Guth, provides compelling solutions to problems like horizon and flatness through a period of rapid exponential expansion \cite{Guth:1980zm}, it does not address the persistent issue of the initial singularity. This limitation has motivated extensive research into alternative cosmological paradigms, particularly bouncing cosmologies \cite{Wang:2003yr, Novello:2008ra, Novello:2008ra, deHaro:2012xj, Battefeld:2014uga, Ijjas:2016tpn, Brandenberger:2012zb}. In such scenarios, the universe undergoes a contracting phase until it reaches a finite minimum radius, followed by a smooth transition to expansion. This cyclic behavior inherently avoids the singularity problem, as the universe never collapses to a point of infinite density. Instead, the bounce marks a transition from contraction to expansion, offering a potentially singularity-free description of cosmic evolution \cite{Battye:2006mb, Cai:2014bea}.

Bouncing cosmologies have attracted significant attention within various modified gravity frameworks, including ekpyrotic models \cite{Khoury:2001wf}, matter bounces \cite{Cai:2012va}, and loop quantum cosmology \cite{Ashtekar:2006wn}. These approaches not only resolve the singularity problem but also provide testable predictions for cosmological observables, such as perturbations in the cosmic microwave background and the large-scale structure of the Universe \cite{Brandenberger:2016vhg}. Recent advances in gravitational theory and observational cosmology continue to refine these models, making them a vibrant area of research in modern theoretical physics.
 
A successful bouncing cosmological model must satisfy four fundamental conditions \cite{Cai:2007qw, Cai:2008ed}:

(i) Scale factor dynamics: The universe undergoes contraction ($\dot{a}(t)<0$) followed by expansion ($\dot{a}(t)>0$), with a smooth transition at the bounce.\\
(ii) Hubble parameter evolution: Correspondingly, $H<0$ during contraction, $H>0$ during expansion, with $H=0$ precisely at the bounce point.\\
(iii) Energy condition violation: Near the bounce, $\dot{H} = -4\pi G\rho(1+\omega) > 0$ indicates a necessary violation of the null energy condition.\\
(iv) Equation of state behaviour: The EoS parameter $\omega$ must cross the phantom divide ($\omega=-1$) in the bounce vicinity, exhibiting quintom-like behaviour.

In recent years, significant advances have been made in bouncing cosmology within various modified gravity frameworks. Four prominent approaches have emerged: (i) $f(R)$ gravity, where successful bouncing scenarios have been developed through both vacuum \cite{Bamba:2008ut} and matter-coupled solutions \cite{Odintsov:2014gea, Odintsov:2015zza}; (ii) $f(R,G)$ gravity incorporating Gauss-Bonnet terms \cite{Singh:2022gln, Shaily:2024rjq}; (iii) $f(R,T)$ gravity with matter-geometry coupling \cite{Singh:2018xjv, Singh:2022jue, Singh:2023gxd}; and (iv) $f(Q,T)$ hybrid theories \cite{Agrawal:2021rur, Lalke:2023cia, Singh:2024tur}. 

\subsection{\textbf{Relation to Other Modified Gravity Theories}}

It is true that in the coincidence gauge – which we effectively use by working in the FLRW metric with a specific connection – the gravitational field equations of many $f(Q) $ models can exhibit a mathematical structure similar to those of $f(T)$ theories, as both are built from first derivatives of the metric. However, the novelty and physical content of our model are not compromised for the following reasons:
\begin{enumerate}
    \item Different Fundamental Action: Our action (Eq. \ref{1}) is not the $f(T)$ action. It explicitly includes a massive Proca field ($w_i$) for the Weyl vector with its own kinetic term ($- \frac{1}{4} W_{\mu\nu} W^{\mu\nu}$) and mass term. This is an additional fundamental degree of freedom not present in standard $f(T)$ gravity.
    \item Different Constraint Structure: Our action includes a Lagrange multiplier $\lambda $ that enforces the condition $R + 6(\nabla\mu w^\mu - w_\mu w^\mu) = 0$, embedding the theory within a specific Weyl integrable geometry. This imposes a different constraint on the geometry compared to the teleparallel framework of $f(T)$.
   \item Novel Phenomenological Mechanism: The interplay between the power-law $f(Q) = \alpha Q^{\xi}$, the mass m of the Weyl vector, and the Lagrange multiplier $\lambda$ creates a unique geometric mechanism for NEC violation and bounce. Our analysis in Section III.A and III.B detail how these specific parameters control the bounce energy scale and the duration of instability, a study not performed in the $ f(T) $ context.
\end{enumerate}
Therefore, while the effective gravitational dynamics for the scale factor might share similarities with some $f(T)$ models, the underlying theory, its degrees of freedom, and the microphysics of the bounce mechanism are distinct. Our work is the first to explore this specific bouncing cosmology within the Weyl-type $ f(Q)$ paradigm.

The literature on quintom cosmology in $f(T)$ or equivalent modified gravity theories is active, focusing on analyzing these models to explain the universe's expansion. The pivotal research areas include dynamical system analysis to study the stability and consistency of the models, by investigating their capability to express phantom-like and quintessence-like functioning, and demonstrating how these models correspond to recent observational data, such as the DESI dataset. Many studies use dynamical system techniques to analyze the stability of the critical points in $f(T)$ and similar models \cite{Capozziello:2011hj, Cai:2015emx, Paliathanasis:2016vsw, Duchaniya:2022rqu}. This helps determine if these models can achieve a stable, accelerating late-time cosmology, a necessary condition for them to be considered viable explanations for the universe's expansion.

A consequential tendency is showing the dynamical equivalence between modified gravity models and quintom models, suggesting these theories can provide a unified framework for dark energy. A major finding is that many modified gravity theories, such as $f(Q)$ gravity, are dynamically equivalent to a quintom-like scenario. This means the complex dynamics of modified gravity can be mapped onto a simpler, two-scalar-field quintom model, providing a powerful theoretical tool for analysis. Cosmological data comparison: Researchers are actively comparing theoretical predictions from quintom cosmology in modified gravity with new cosmological data, particularly from the Dark Energy Spectroscopic Instrument (DESI). Recent findings suggest some models are consistent with or even favored by this data over the standard Lambda-CDM model.

Several studies in the literature examine the specific behavior of these models, such as their ability to have the dark energy equation of state cross the phantom divide line $(\omega = -1)$. This crossing is an important criterion of the quintom scenario, and the work is focused on understanding the conditions under which it occurs and the stability of the system during the crossing of the phantom divide line.

Some of the works have also been focused on symmetric teleparallel gravity ($f(Q)$), where the non-metricity tensor $Q$ replaces curvature as the fundamental geometric object. Cruz-Dombriz et al. have discussed cosmological bouncing solutions in extended teleparallel theories of gravity \cite{delaCruz-Dombriz:2018nvt}. Several innovative approaches have been developed in this framework: The order reduction method for field equations \cite{Bajardi:2020fxh}, combination with loop quantum cosmology principles \cite{Agrawal:2022vdg}, Lagrangian reconstruction techniques \cite{Gadbail:2023loj}, anisotropic Bianchi-I models \cite{Sharif:2024bwy}, comprehensive stability analyses \cite{Koussour:2024wtt} and Weyl-type extensions with matter coupling \cite{Zhadyranova:2024hbc}. Basilakos et al. \cite{Basilakos:2025olm} examine that the combination of two well-behaved fields, that is, a scalar field arising from the non-linear $f(Q)$ form, and the scalar field associated with the non-trivial connection, appears effectively as a phantom field. We present a novel bouncing cosmology in Weyl-type $f(Q)$ gravity based on these developments. The key geometric feature is the nonmetricity tensor $ Q_{\alpha ij} \equiv \tilde{\nabla}\alpha g_{ij} = 2w_\alpha g_{ij}$, where $ w_\alpha $ is the Weyl vector that preserves scale covariance while introducing new degrees of freedom \cite{Goswami:2023knh}. This framework naturally incorporates the symmetric teleparallel structure and Weyl geometry, offering new possibilities for singularity resolution.

This work provides the first detailed analysis of a bouncing cosmology within the Weyl-type $f(Q)$ framework with a massive vector field, revealing a new geometric mechanism for singularity resolution. Section \ref{sec-2} presents the fundamental theoretical formulation, where we derive the complete set of modified Einstein field equations and establish the dynamical equations governing the system, with particular attention to the role of the non-metricity tensor Q and its relation to the Weyl vector in the FLRW spacetime. Building upon this foundation, Section \ref{sec-3} conducts a thorough investigation of the bouncing scenario, examining the essential kinematic requirements, the evolution of cosmological parameters, energy condition violations near the bounce point, stability criteria through adiabatic perturbations, and the quintom behaviour emerging from scalar field reconstruction. The Concluding Section \ref{sec-4} synthesises our findings, demonstrating how the model successfully satisfies all bouncing criteria while resolving the initial singularity, and discusses the broader implications for early universe cosmology within this modified gravitational framework.

\section{Formulation of Field Equations in Weyl-type $ f(Q) $ gravity }{\label{sec-2}}

The gravitational action for Weyl-type $f(Q)$ gravity can be formulated as \cite{Goswami:2023knh,BeltranJimenez:2017tkd,Heisenberg:2023lru}:

\begin{equation}\label{1}
S = \int \bigg[ \kappa^2 f(Q) - \frac{1}{4} W_{\mu\nu} W^{\mu\nu} - \frac{1}{2} m^2 w_\mu w^\mu + \lambda (R + 6\nabla_\mu w^\mu - 6w_\mu w^\mu) + \mathcal{L}_m \bigg] \sqrt{-g} d^4x,
\end{equation}

where $ \kappa^2 \equiv (16\pi G)^{-1} $ represents the gravitational coupling constant \cite{Will:2014kxa}. The action incorporates several fundamental components: The function $f(Q)$ generalizes the non-metricity scalar $Q = -6w_\mu w^\mu$ in symmetric teleparallel gravity \cite{BeltranJimenez:2019tme}, while the Weyl vector field $w_\mu$ with mass $m$ appears through both its field strength tensor $W_{\mu\nu} = \nabla_\nu w_\mu - \nabla_\mu w_\nu$ \cite{Blagojevic:2002du} and a Proca-type mass term \cite{Proca:1936fbw}. The matter sector is described by $\mathcal{L}_m$ following standard formulations \cite{Weinberg:1972kfs}, and the Lagrange multiplier $\lambda$ enforces the condition of vanishing total curvature $R + 6(\nabla\mu w^\mu - w_\mu w^\mu) = 0$ that characterizes Weyl geometry \cite{Scholz:2017pfo}.

Riemannian geometry uses a metric and a unique, torsion-free, metric-compatible connection (Levi-Civita), defining curvature through length and angle changes; Metric-Affine Geometry (MAG) treats the metric and a general linear connection (with potential torsion and non-metricity) as independent fields, allowing more complex theories where gravity involves energy-momentum and intrinsic spin (hypermomentum), generalizing Einstein's GR to include torsion and non-metricity beyond just spacetime curvature. The non-metricity imparts a geometric source for gravity, which is used as a new effective stress-energy originating from the geometric source, instead of conventional matter, and possibly simulating dark matter or dark energy. 

The physical role of the Weyl vector is basically a generalization of Riemannian geometry that extends gravity to include local scale invariance in Weyl geometry. It can be interpreted as a purely geometric field, representing the curvature of spacetime that is not considered by the Ricci tensor, which explains the part of gravity that is equivalent to Einstein general relativity \cite{Berezin:2022odj}.

In particular, it is relevant in theories like metric-affine gravity and symmetric teleparallel gravity, where a non-metric connection, instead of the metric, determines the geometry, and a derived quantity known as the non-metricity tensor introduces a modification in the gravity. This geometric effect can be shown as an effective energy-momentum tensor, affecting the dynamics of the universe without requiring new matter fields. The distinction between metric-affine and Riemannian pictures is that Riemannian geometry uses a metric $g$ to define the curvature, whereas in metric-affine geometry, the metric $ g $ and the connection $\Gamma $ are used as independent fields, which allows for non-zero torsion and non-metricity. The Riemannian picture is a particular case of the more general metric-affine picture, where the connection is metric-compatible and torsion-free.

The  degrees of freedom of Weyl type $f(Q)$ gravity can be discussed in the following manner as:\\
(i) the $f(Q)$ gravity propagates two tensor gravitational wave modes, which are the plus and cross polarizations like General Relativity, \\
(ii) The primary aspect is that the scalar degree of freedom, which is found in $f(R)$ gravity theories, is absent in the linearized $f(Q)$ gravity because the first-order perturbation of the connection does not lead to a modification of the linearized field equations in vacuum. \\
(iii) The scalar modes disappear due to gauge invariance, which is the property of a theory being able to be transformed without changing its physics. This means that the scalar modes are not physical, or propagating, degrees of freedom.\\
(iv) In the linearized limit, the field equations of Weyl-type $f(Q)$ gravity with specific assumptions are the same as those of GR, which shows that the number of propagating degrees of freedom is the same, i.e., two tensor modes.

The redundancy of degrees of freedom is addressed by first identifying the redundant links or joints and then accounting for them in calculations, such as with the Kutzbach criterion, by subtracting the redundant degrees from the total. For systems with redundancy, like robotic arms or the human body, control strategies are developed to exploit the extra freedom to achieve a task by satisfying additional constraints like obstacle avoidance or increased dexterity. 

The Weyl-type $f(Q)$ gravity is not directly equivalent to $f(T)$ gravity, but they are related through the specific conditions of the coincidence gauge in modified gravity theories. In this gauge, the non-metricity scalar $ Q $ is often assumed to vanish, and the connection is fixed to be torsionless and symmetric. The field equations of Weyl-type $f(Q)$ gravity, under the conditions of the coincidence gauge, become equivalent to the field equations of $f(T)$ gravity. This is because the choice of gauge converts the gravitational description from a non-metricity-dominated one to a torsion-dominated one.

The coincidence gauge in symmetric teleparallel gravity (STEGR) is a particular choice of coordinates where the affine connection vanishes, and simplifies the geometry so that covariant derivatives become partial derivatives. This differs fundamentally from the teleparallel equivalence of general relativity (TEGR) and related theories, such as $f(T)$ gravity, which employ a distinct framework with non-zero torsion. The coincidence gauge works by removing the independent connection entirely from the equations, which is appropriate for STEGR's zero-curvature and zero-torsion geometry and is non-degenerate in this context. However, ignoring correspondence is not an issue. Alternatively, it shows a different but mathematically equivalent formulation of gravity.

The coincidence gauge affects the generality and uniqueness of solutions in Weyl-type $f(Q)$ gravity by changing the number of physical degrees of freedom. Fixing the gauge condition breaks the smooth transformation symmetry, which in turn changes the dynamics and can lead to a specific set of solutions, potentially reducing the overall generality of the solutions compared to the non-gauged theory.

In standard $f(Q)$ gravity, the connection is often assumed to be coincident with the Levi-Civita connection $(\Gamma _{ij }^{k }=\Gamma _{ij }^{k }|_{LC})$, which is equivalent to setting the non-metricity to zero. Non-coincident $f(Q)$ gravity abandons this assumption, allowing the connection to be independent of the metric and introducing non-metricity $(Q\ne 0)$ and torsion. In $f(Q)$ gravity with non-coincident connections, the Einstein field equations can behave like a two-scalar-field quintom model, where the Phantom field driving the late-time acceleration is not an added scalar field but an effective description arising from the non-trivial connection. This happens because the non-zero, non-coincident connection introduces an extra degree of freedom that couples with the non-metricity scalar $ Q $  and standard matter, leading to an effective dark energy with properties similar to a quintom fluid.

The motivation for modified gravity is to address cosmological mysteries like dark energy and dark matter, including the Bouncing scenario by changing the theory of general relativity at large scales, whereas the motivation for quantum fields is to explain the behavior of matter and the energy density at the minor scales and to unify all fundamental forces. Both approaches are ultimately driven by the need for a more complete and consistent understanding of the universe, particularly by attempting to reconcile gravity with the quantum mechanics that regulate the other forces and by analyzing phenomena like black holes and the early universe.

This action combines elements from several well-established theoretical frameworks: The $f(Q)$ term extends symmetric teleparallel gravity \cite{BeltranJimenez:2019acz}. At the same time, the Weyl sector incorporates both gauge-invariant kinetic terms and explicit mass contributions. The geometric constraint maintains the essential features of Weyl geometry while allowing for modified gravitational dynamics. Such constructions have been shown to provide ghost-free alternatives to traditional modified gravity theories \cite{Lin:2021uqa}, while maintaining consistency with cosmological observations \cite{DAmbrosio:2021zpm}. The resulting theory offers a promising framework for investigating both early universe cosmology and late-time acceleration \cite{Goswami:2023knh}.
%The brief introduction to Weyl geometry, which introduces the intrinsic vector field $ w_{i} $, non-metricity scalar $ Q $, and the tensor $ W^{ij} $ is described in the Appendix.
 
The non-metricity scalar $ Q $ is defined as,
\begin{equation}{\label{2}}
    Q \equiv - g^{i j} \bigg( {L^\alpha}_{\beta i}  {L^\beta}_{j \alpha} - {L^\alpha}_{\beta \alpha} {L^\beta}_{i j} \bigg).
\end{equation} 
where $  {L^\alpha}_{i j} $ denotes the deformation tensor which is defined as
\begin{equation}{\label{3}}
   {L^\alpha}_{i j} = - \frac{1}{2} g^{\alpha  \gamma} \bigg( Q_{i \gamma j} + Q_{j \gamma i} - Q_{\gamma i j} \bigg).
\end{equation}
The covariant derivative of the metric tensor in the Riemannian geometry is zero, i.e., $ \nabla_\alpha g_{ij} =0 $. In Weyl geometry, we have \cite{Haghani:2012bt}
\begin{equation}{\label{4}}
     Q_{\alpha i j} \equiv  \tilde{\nabla}_\alpha g_{i j} = \partial_\alpha g_{i j} - {\tilde{\Gamma}^\eta}_{\alpha i} g_{\eta j} - {\tilde{\Gamma}^\eta}_{\alpha j} g _{\eta i} = 2 w_\alpha g _{i j}.
\end{equation}
where $ \Gamma^\alpha_{i j} $ denotes the Christoffel symbol with respect to the metric $ g_{i j} $ and a semi-metric connection $ {\tilde{\Gamma}^\alpha}_{i j} $ which is defined as
\begin{equation}{\label{5}}
    {\tilde{\Gamma}^\alpha}_{i j} \equiv \Gamma^\alpha_{i j} + g_{i j} w^\alpha - \delta^\alpha_i w_j - \delta^\alpha_j w_i  
\end{equation}
From Eqs. (\ref{2}-\ref{4}), we get a relation,
\begin{equation}{\label{6}}
      Q = - 6 w^2.
\end{equation}

The corresponding Proca-type equation is derived by applying variation to the action (\ref{1}) concerning the vector field $ w $,
\begin{equation}{\label{7}}
     \nabla^j W_{i j} - (m^2 + 12 k^2 f_Q + 12 \lambda) w_i = 6 \nabla_i \lambda.
\end{equation}
Comparing Eq. (\ref{7}) with the standard Proca equation, we define the effective dynamical mass of the vector field $ w $ as
\begin{equation}{\label{8}}
     m^2_{eff} = m^2 + 12 k^2 f_Q + 12 \lambda.
\end{equation}
Taking variation on the action (\ref{1}) with respect to the metric tensor $ g_{ij} $, we have the field equation as
\begin{multline}{\label{9}}
 \frac{1}{2} (T_{i j} + S_{i j} ) = - \frac{k^2}{2} g_{i j}  f -6 k^2 f_Q w_{i} w_{j} + \lambda ( R_{i j} - 6 w_{i} w_{j} + 3 g_{i j} \nabla_\eta w^{\eta})+ \\
  3 g_{i j} w ^{\eta} \nabla_{\eta} \lambda - 6 w_{(i \nabla_j)} \lambda + g_{i j}  \square \lambda - \nabla_{i} \nabla_{j} \lambda,
\end{multline}
where $ f_Q $ denotes the first derivative of $ f $ with respect to $ Q $, and $ T_{ij} $ and $ S_{ij} $ represent the energy-momentum tensor of the universe's content and the rescaled energy-momentum tensor of the free Proca field
\begin{align}
    T_{i j}  \equiv & -\frac{2}{\sqrt{-g}} \frac{\delta(\sqrt{-g} L_m)}{\delta g^{i j}}, \label{10}\\
    S_{i j} = &-\frac{1}{4} g_{i j}  W_{\eta \alpha} W^{\eta \alpha} + W_{i \eta} W_{j}^{\eta} -\frac{1}{2} m^2 g_{i j} w_{\eta} w^{\eta} + m^2 w_{i} w_{j}, \label{11}
\end{align}
respectively.
%%%%%%%%%%%%%%%%%%%%%%%%%%%%%%%%%%%%%%%%%%%%%%%%%%%%%%%%

We consider the spatially flat Friedmann-Lema$\hat{i}$tre-Robertson-Walker (FLRW) metric, which characterizes the cosmological evolution within a flat geometric framework, to conduct a more profound analysis for an in-depth comprehension of bouncing Weyl-type gravity. 
\begin{equation}\label{12}
      ds^2 = -N^2(t)dt^2 + a^2(t)\delta_{ij}dx^idx^j ,
 \end{equation}
where $a(t)$ is the scale factor, $ N(t)$ is the lapse function, and $ N(t)=1 $ in the standard case.

The vector field $ w_{i} $ is considered as $  w_{i} = [0,0,0,\psi(t)] $. Therefore, $ w^2 = w_i w^{i} = - \psi^2(t) $ and $ Q = -6 w^2 = 6 \psi^2(t) $. The Lagrangian of the perfect fluid is considered as $ L_m = p $ where $ p $ denotes the pressure of the perfect fluid. Therefore,

\begin{equation}{\label{13}}
    T^i_j = (p + \rho)u^i u_j + p \delta^i_j = diag( p, p, p, -\rho),
\end{equation}
where $ \rho $ denotes the matter-energy density of the perfect fluid. In a comoving coordinate system, the velocity vector is given by $ u^i = (0,0,0,1) $ such that $ u^i u_i = -1 $. For the metric (\ref{12}), the generic Proca equation can be written as, 
 \begin{align}
    \dot{\psi} &= \dot{H} + 2 H^2 + \psi^2 - 3 H \psi, \label{14}\\
     \dot{\lambda} &= (-\frac{1}{6} m^2 - 2 k^2 f_Q -2 \lambda ) \psi = - \frac{1}{6} m_{eff}^2 \psi,  \label{15} \\
     \partial_i \lambda &= 0.\label{16}
 \end{align}
and the evaluated field equations (\ref{9}) are 
\begin{align}
    \frac{1}{2} \rho &= \frac{k^2}{2} f - \bigg( 6 k^2 f_Q + \frac{1}{4} m^2 \bigg) \psi^2 
      - 3 \lambda (\psi^2 - H^2 ) - 3 \dot{\lambda} (\psi - H),{\label{17}} \\
      -\frac{1}{2} p &= \frac{k^2}{2} f + \frac{m^2 \psi^2}{4} + \lambda (3 \psi^2 + 3 H^2 + 2 \dot{H}) + (3\psi + 2 H ) \dot{\lambda} +\ddot{\lambda}.  {\label{18}}  
\end{align}

To explore the matter bounce in the framework of the Weyl-type $ f(Q) $ gravity, we adopt a power law functional form of $ f(Q) $ as follows: 
\begin{equation}{\label{19}}
    f(Q) = \alpha Q^{\xi},
\end{equation}  
where $ \alpha $ and $ \xi $ are designated as model parameters. The parameter $ M^2 = \frac{m^2}{k^2} $ represents the mass of the Weyl field. 
Using Eqs. (\ref{17}), (\ref{18}) and (\ref{19}), we obtain
\begin{equation}{\label{20}}
   \frac{1}{2} \rho = \frac{1}{2} (\alpha Q^{\xi}) - \bigg( 6 (\alpha \xi Q^{\xi-1}) + \frac{1}{4} m^2 \bigg) \psi^2 - 3  (\psi^2 - H^2 ),
\end{equation} 
and
\begin{equation}{\label{21}}
    -\frac{1}{2} p = \frac{1}{2}(\alpha Q^{\xi}) + \frac{m^2 \psi^2}{4} +  (3 \psi^2 + 3 H^2 + 2 \dot{H}), 
\end{equation}
where we take $ \lambda = k^2 = 1 $ without loss of generality. For further calculation, we consider an ansatz $ \psi(t) = H(t) $. This ansatz allows us to isolate the cosmological effects of the non-metricity scalar Q coupled to a time-like Weyl vector proportional to the expansion rate. The identification $\psi(t) = H(t)$ is a consistent simplifying ansatz chosen to explore a specific class of solutions within Weyl-type $f(Q)$ gravity, not a general outcome of the theory. Its justification is two-fold:
\begin{enumerate}
    \item Mathematical Consistency: From the vector field equation (14), $\dot{\psi} = \dot{H} + 2 H^2 + \psi^2 - 3 H \psi$. The ansatz $ \psi = H $ is a fixed point of this equation, as substitution yields an identity ($0 = 0$). This makes it a consistent and tractable choice that reduces the system’s complexity.
   \item Physical Motivation: In a homogeneous and isotropic universe (FLRW metric), if the Weyl vector is assumed to be aligned with the cosmic flow (i.e.,  $  w_{i} = [0,0,0,\psi(t)] $]), it is natural to explore the case where its magnitude is tied to the fundamental rate of expansion, $H(t)$. This ansatz effectively models a scenario where the Weyl geometric distortion scales with the Hubble flow.
\end{enumerate}
We explicitly acknowledge that this choice selects a particular subset of solutions. It restricts the independent degrees of freedom in the connection by linking the Weyl vector directly to the metric’s first derivative. Distinct choices for $\psi(t)$ would indeed lead to different dynamical evolution, potentially without a clear bounce. The purpose of this work is to demonstrate the existence of a viable bouncing cosmology within this theory under a well-motivated, simplifying condition. This ansatz allows us to isolate the cosmological effects of the non-metricity scalar $ Q $ coupled to a time-like Weyl vector proportional to the expansion.

The isotropic pressure $ p $ and the energy density $ \rho $ in terms of $ H $ are given by
\begin{equation}{\label{22}}
     \rho(t) = - \frac{1}{2} m^2 H(t)^2 - \alpha 6^{\xi}  (2 \xi - 1) (H(t)^2)^{\xi}
\end{equation} 
\begin{equation}{\label{23}}
     p(t) = - \frac{1}{2} (24 + m^2) {H(t)}^2 - \alpha 6^{\xi}   {H(t)^2}^{\xi} - 4 H'(t)
\end{equation}
The EoS parameter $ \omega $ is obtained as
\begin{equation}{\label{24}}
     \omega(t) =  \frac{( 24+ m^2 ) {H(t)}^2 + 2 ( \alpha \hskip0.02in 6^{\xi} ({H(t)}^2)^{\xi}) + 4 H'(t)}{ m^2 {H(t)}^2 + \alpha \hskip0.02in 3^{\xi}\hskip0.01in (2)^{1+ \xi}  (-1+ 2 \xi) ({H(t)}^2)^{\xi}}
\end{equation}

In the following analysis, we examine the bouncing cosmology scenario within the Weyl-type $f(Q)$ gravity framework, focusing on its distinctive dynamical characteristics. This paradigm offers an alternative to conventional Big Bang cosmology by proposing a cyclical evolution consisting of three phases: a contracting phase ($\dot{a}(t)<0$), a non-singular bounce occurring at finite minimum volume, and subsequent expansion ($\dot{a}(t)>0$). Such models provide a compelling resolution to the initial singularity problem while maintaining compatibility with observational constraints \cite{Brandenberger:2012zb}.

\section{ Physical features of the model}{\label{sec-3}}

The successful cosmological bounce requires several crucial characteristics to be discussed. First and foremost, the null energy condition (NEC) must be violated in the vicinity of the bounce point, as evidenced by $\dot{H} > 0$ during this transitional period \cite{Cai:2007qw}. The geometric origin of Null Energy Condition (NEC) violation in physics relates to how spacetime curvature interacts with energy, essentially meaning gravity can act repulsive for light (null geodesics) instead of purely attractive, allowing light beams to diverge (Raychaudhuri equation) or matter to have negative pressure i.e., dark energy causing accelerated expansion, often seen in modified gravity or quantum scenarios like tunneling between vacua, breaking classical GR expectations. 

This NEC violation enables the necessary transition from contraction to expansion. Additionally, the equation of state parameter $\omega$ must exhibit quintom-like behavior, crossing the phantom divide ($\omega = -1$) during the bounce evolution \cite{Cai:2008ed}. This characteristic transition from $\omega < -1$ (phantom regime) to $\omega > -1$ (quintessence regime) serves the dual purpose of preventing a future Big Rip singularity while facilitating the bounce mechanism. Furthermore, all physical quantities, including the energy density $\rho$, pressure $p$, and curvature invariants, must remain finite throughout the entire bounce process to ensure regularity \cite{Bamba:2008ut, Novello:2008ra}.

Within the Weyl-type $f(Q)$ gravity framework, these conditions naturally emerge through several distinctive features. The geometric effects mediated by the non-metricity scalar $Q$ and its associated Weyl vector $w_\mu$ modify the standard Friedmann equations in a manner that permits controlled NEC violation. Moreover, the theory supports stable transitions between cosmological phases while maintaining physical consistency. Our subsequent investigation will quantitatively examine these aspects through detailed analysis of key parameters: the scale factor evolution $a(t)$, Hubble parameter $H(t)$ behavior, deceleration parameter $q(t)$ dynamics, energy condition violation patterns, and stability criteria via the sound speed $C_s^2$ and adiabatic index $\Gamma$. This comprehensive approach demonstrates how Weyl-type $f(Q)$ gravity robustly satisfies all theoretical requirements for a viable bouncing cosmology, while offering novel insights into early universe dynamics. Since Eqs. (\ref{17}) and (\ref{18}) are highly nonlinear and nonhomogeneous. Therefore, we employ the reconstruction method, assuming a bounce-compatible scale factor ansatz:

\qquad Following the standard reconstruction approach in cosmology, we impose a specific form for the scale factor, $a(t)$, which is known to satisfy the basic kinematic conditions for a nonsingular bounce ($a > 0$, $ \dot{a}(0) = 0$, $\ddot{a}(0) > 0$). This ansatz is not an ``exact solution” derived a priori from the field equations. Instead, it is a physically motivated input that allows us to solve the highly nonlinear modified Friedmann equations (17) and (18) and test whether the Weyl-type $f(Q)$ gravity framework can support such a bouncing history. The success of this reconstruction demonstrates the compatibility of the theory with a bouncing universe. Therefore, to solve these equations, we parameterize the scale factor $ a $ in terms of cosmic time $ t $ as \cite{Odintsov:2016tar, Saridakis:2018fth}:
\begin{equation}{\label{25}}
    a(t) = (\beta \sinh^{2}t + \gamma )^{n/3},
\end{equation}
where $ \beta $, $ \gamma $, and $ n $ are arbitrary positive constants, and we consider the first term in the expansion of $ \sinh(t) $ to avoid the complexity of the computation.  

\begin{table}[htbp]
\caption{ Evolution of the universe}
%\begin{ruledtabular}
\begin{center}
\label{tabparm1}
\renewcommand{\arraystretch}{1.5}
\setlength{\tabcolsep}{14pt}
\begin{tabular}{lc c c c r} 
\hline\hline
\quad & $ Scale factor $ & $ q $ & $ H $ &  \textit{Time interval} & \qquad \textit{Evolution of Universe}
\\
\hline 
\quad & $ a>0 $ (decreases steadily)& $ <0 $ & $ <0 $ &  $ -1.61< t < 0 $  &  $ q \ominus H $\footnote{ Accelerated contraction for $ n=0.35 $, $ \gamma=1 $}
\\
\quad & $ a>0 $ (decreases steadily)& $ <0 $ & $ <0 $ &  $ -2.44< t < 0 $  &  $ q \ominus H $\footnote{ Accelerated contraction for $ n=0.50 $, $ \gamma=2 $}
\\
\quad & $ a>0 $ (decreases steadily) & $ <0 $ & $ <0 $ &  $ -3.44< t < 0 $  &  $ q \ominus H $\footnote{ Accelerated contraction for $ n=0.75 $, $ \gamma=3$}
\\
\hline 
\quad & $ a_{min}>0 $\footnote{$ a_{min}\approx 1 $ for $ n=0.35 $, $ \gamma=1 $; $ a_{min}\approx 1.122 $ for $ n=0.50 $, $ \gamma=2 $; $ a_{min}\approx 1.316 $ for $ n=0.75 $, $ \gamma=3 $} & $ <0 $ & $ =0 $ & $ ~~t \approx 0 $ &   $ q_{max}\odot H $\footnote{ Bounce with maximum acceleration $ \forall n,\gamma $} 
\\
\hline 
\quad & $ a>0 $ (increases steadily) & $ <0 $ & $ <0 $ &  ~~$ 0< t < 1.61 $  &  $ q \oplus H $\footnote{ Accelerated expansion for $ n=0.35 $, $ \gamma=1 $}
\\
\quad & $ a>0 $ (increases steadily) & $ <0 $ & $ <0 $ &  ~~$ 0< t < 2.44 $  &  $ q \oplus H $\footnote{  Accelerated expansion for $ n=0.50 $, $ \gamma=2 $}
\\
\quad & $ a>0 $ (increases steadily) & $ <0 $ & $ <0 $ &  ~~$ 0< t < 3.44 $  &  $ q \oplus H $\footnote{  Accelerated expansion for $ n=0.75 $, $ \gamma=3 $}
\\
\hline 
\\
\end{tabular}    
%\end{ruledtabular} 
\end{center}
\end{table}
The Hubble parameter is evaluated as 
\begin{equation}{\label{26}}
    H(t) = \frac{\dot{a}}{a} = \frac{2 n t \beta}{3 (t^2 \beta +\gamma)}
\end{equation}

The deceleration parameter in terms of cosmic time $ t $ using equation (\ref{26}) is determined as 
\begin{equation}{\label{27}}
    q(t) = -1 - \frac{\dot{H}}{H^2} = \frac{3 t^2 \beta - 2 n t^2 \beta - 3 \gamma}{2 n t^2 \beta}
\end{equation}

\begin{figure}
\begin{center}
     \subfloat[]{\label{a(t)} \includegraphics[scale=0.43]{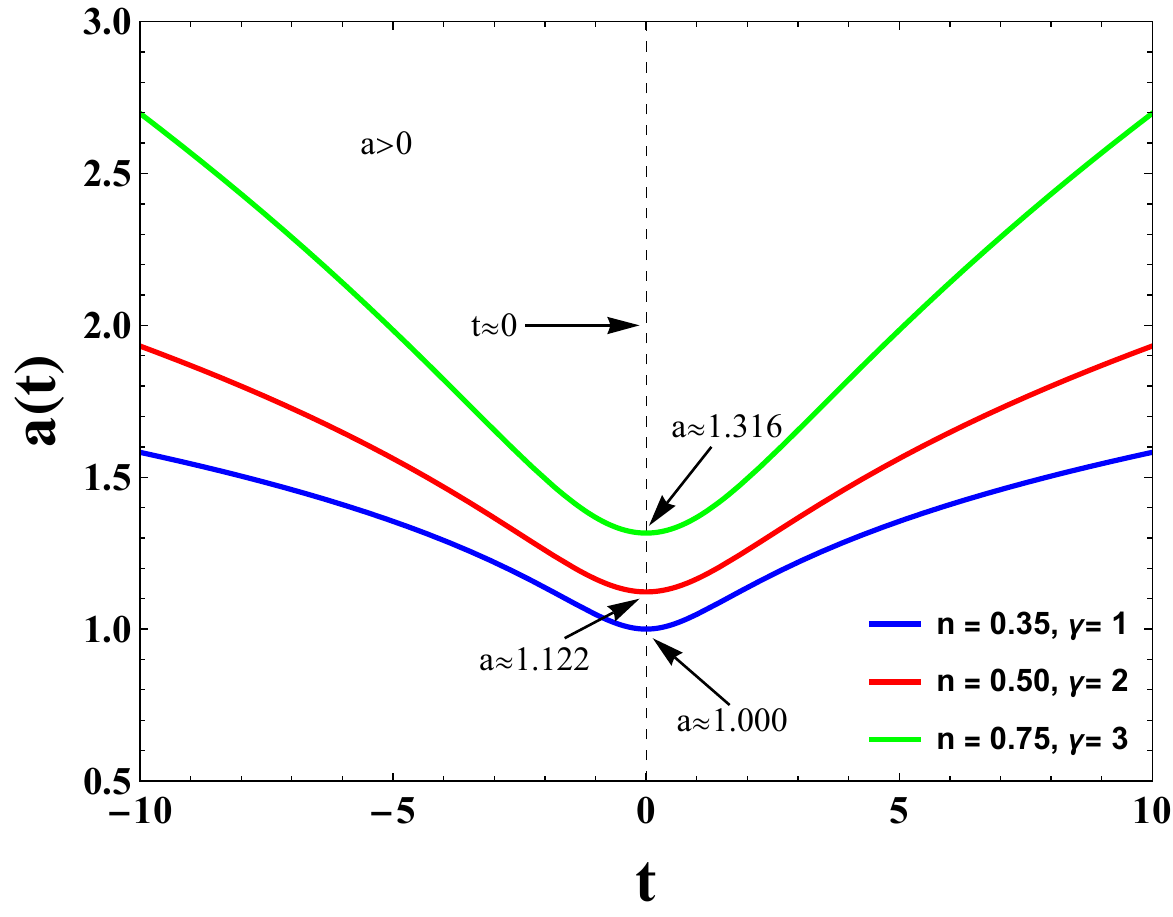}}\hfill
     \subfloat[]{\label{H(t)} \includegraphics[scale=0.43]{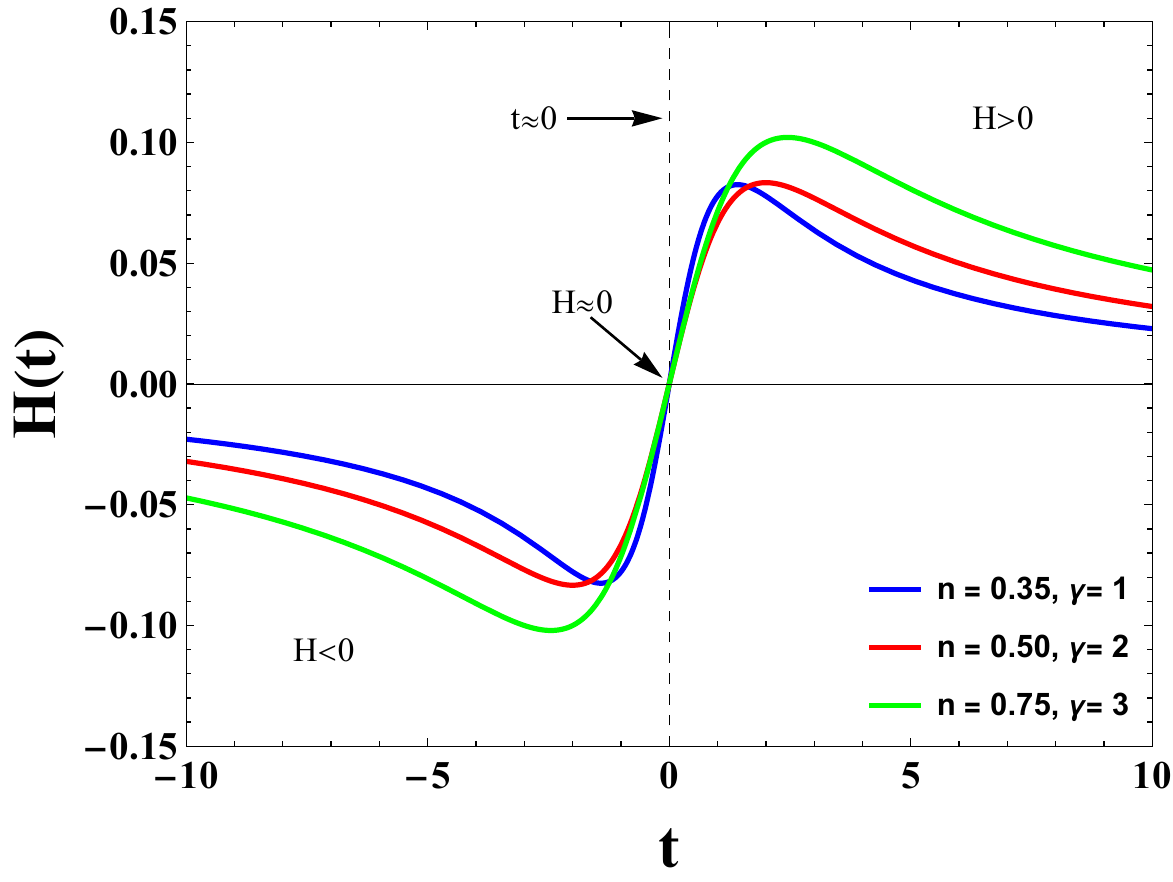}}\hfill
      \subfloat[]{\label{q(t)} \includegraphics[scale=0.43]{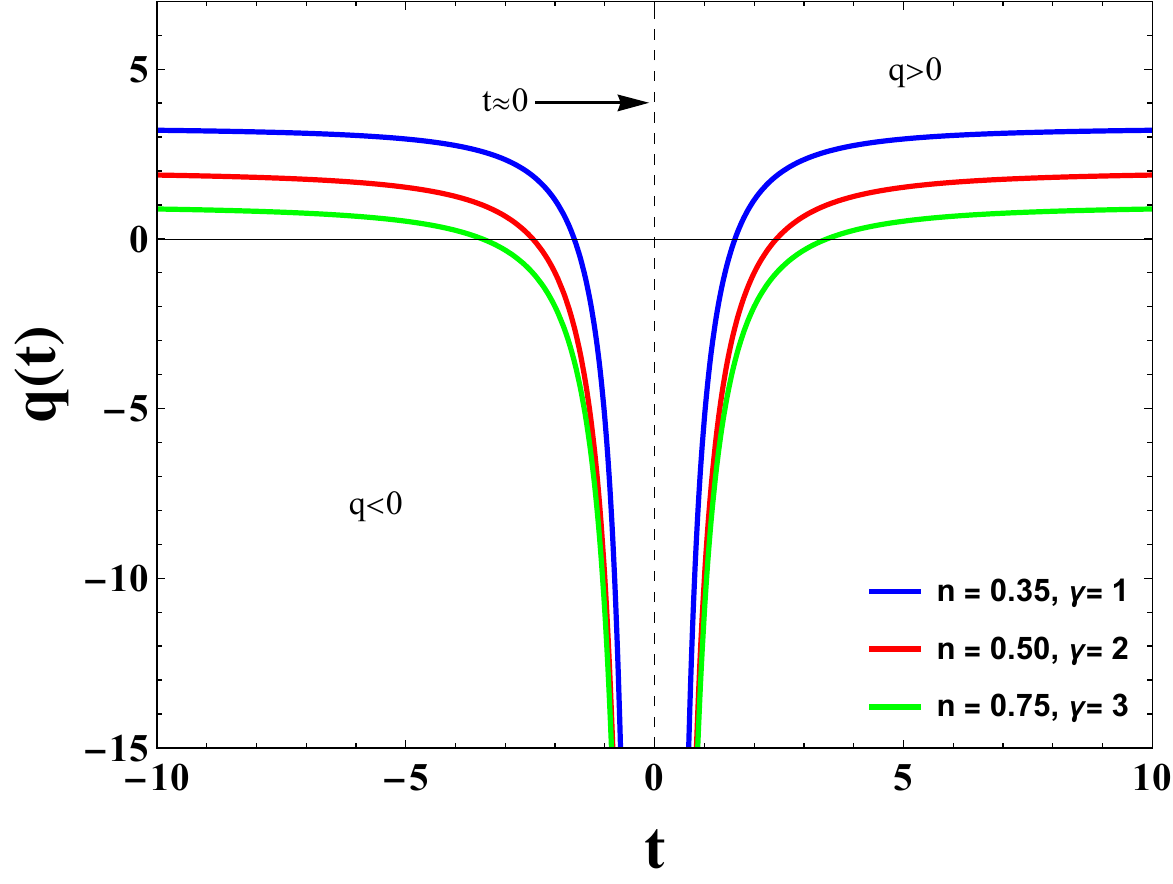}} \hfill
      \subfloat[]{\label{Hr(t)} \includegraphics[scale=0.43]{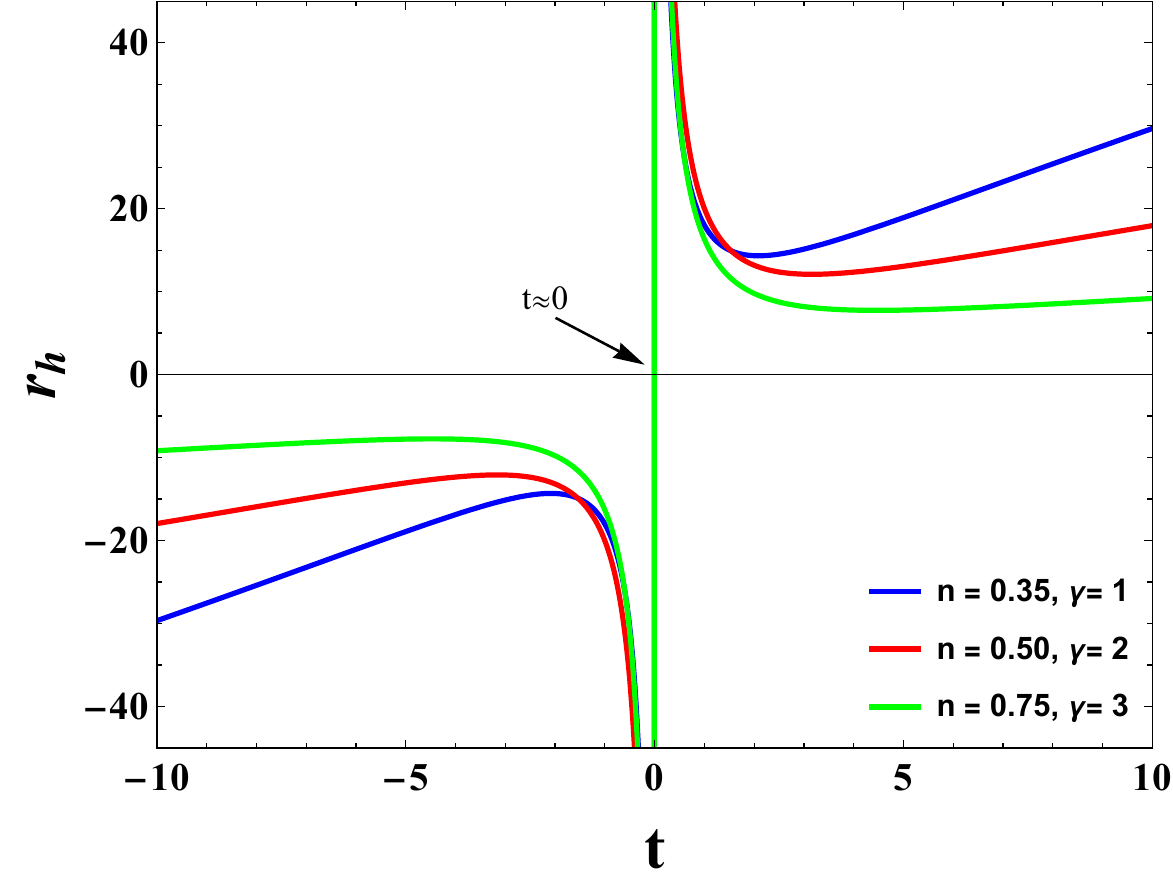}}
\end{center}
\caption{ The variation of $ a $, $ H $, $ q $ and the Hubble radius $ r_h $ for distinct model parameters.}
\label{fig.1}
\end{figure}

The trajectories of the physical parameters are studied on the model parameters $ \beta =0.3 $, $ n=0.35 $, $ \gamma=1 $; $ \beta =0.3 $, $ n=0.50 $, $ \gamma=2 $; and $ \beta =0.3 $, $ n=0.75 $, $ \gamma=3 $. 

As per the observations in Table \ref {tabparm1} and Fig. \ref{fig.1} we see that\\
(i) In Fig. \ref{a(t)}, it is observed that the scale factor $ a(t) < 0$  before bounce, $ a(t)> 0$  after bounce, attains a non-zero minimum value at the bouncing point $ t\approx0$. Thus, we interpret that the model contracts before bounce, expands after bounce, and is nonsingular.\\
(ii) In Fig. \ref{H(t)}, it has been analyzed that the our model contracts ($ H < 0 $) before bounce i.e., contracting collapse, ($ H = 0 $) at the bouncing point $ t\approx0 $, i.e., the bounce transition where $ H $ crosses zero, and the radiation era ($ H > 0 $) after bounce. The peak value of $ |H| $ at the post-bounce phase sets the bounce energy scale, which is denoted as the quasi-highest energy scale. The interval from $ H > 0 $ to the peak value of $ |H| $ at the post-bounce phase is indicated as the Reheating temperature \cite{Lai:2025efh}.\\
(iii) Fig. \ref{q(t)} exhibits the nature of $ q $  in the evolution of the Universe for distinct values of the free model parameters. The deceleration parameter $ q$  transits from a decelerating state $ q > 0$  to an accelerating state $ q< 0$  before bounce, accelerating state to decelerating after bounce, and the model shows extreme acceleration at the bouncing point $ t\approx0$.

The authors may have discussed the distinct forms of development that our universe represents \cite{Singh:2022jue}, including\\
 (a) the super exponential expansion $ q < -1 $;\\
 (b) exponential expansion $-1 \geq q < 0$  ( $ q= -1 $ is also said to be as de-Sitter expansion);\\
 (c) expansion with constant rate $ q=0 $;\\
 (d) accelerating power expansion $ -1 < q < 1 $;\\
 and (e) decelerating expansion $ q>0 $. \\
(iv) In Fig. (\ref{Hr(t)}), the cosmic Hubble radius \cite{Agrawal:2022ppe} $ ( r_h = \frac{1}{aH} ) $ experiences symmetric behavior around the bouncing point.  The Hubble radius $ r_h<0 $ before bounce, $ r_h>0 $ after bounce, and $ r_h=0 $ at the bouncing point ($ t\approx 0 $). This indicates the accelerating behavior of the Universe in later stages.

Under the potential observational implications of a bouncing scenario of the universe, we observe the modified power spectra in the cosmic microwave background (CMB), particularly with oscillations on large scales, and potential deviations from Gaussianity. These effects arise from the contracting phase before the bounce, which can embed unique features on perturbations that are then stretched by inflation and observed in large-scale structures. There are some basic criteria for the potential observations of a bouncing scenario as follows:

(i) The Cosmic Microwave Background: The bounce could embed oscillations or damping on the large-scale fluctuations of the CMB, which differ from the predictions of standard inflation alone. Some bouncing models can predict deviations from a perfectly Gaussian distribution in the CMB temperature variations, though the specific type of non-Gaussianity depends on the bounce model. The potential Observations, especially from the CMB, can be used to place constraints on the parameters of bouncing models and behave as consistency checks for the theory, which can be seen in some studies using WMAP data. 

(ii) Large-scale structures: The large-scale structures in the universe may show a distinct pattern compared to the standard cosmological model due to the effects of the bouncing phase on the distribution of matter and energy. The observational constraints can compare the observed large-scale structures with predictions from bouncing models and can help to constrain the scenario, although the current observational data may not yet have sufficient precision to definitively confirm these effects. 

(iii) Theoretical consistency and other consequences: The bouncing models provide a way to avoid the infinite energy density, infinite temperature, and curvature of the Big Bang singularity, offering a smoother, non-singular beginning to the universe's expansion. The models in the matter bounce scenario are specifically constructed to produce a matter-dominated era in the later expansion phase, which is consistent with observations. Additionally, the well-constructed bouncing model should also provide a natural mechanism for the bounce to transition into the subsequent inflationary epoch and reheating phase, which are crucial to describe the present state of the universe. 

\begin{figure}[h]
\begin{center}
       \subfloat[]{\label{rho(t)} \includegraphics[scale=0.43]{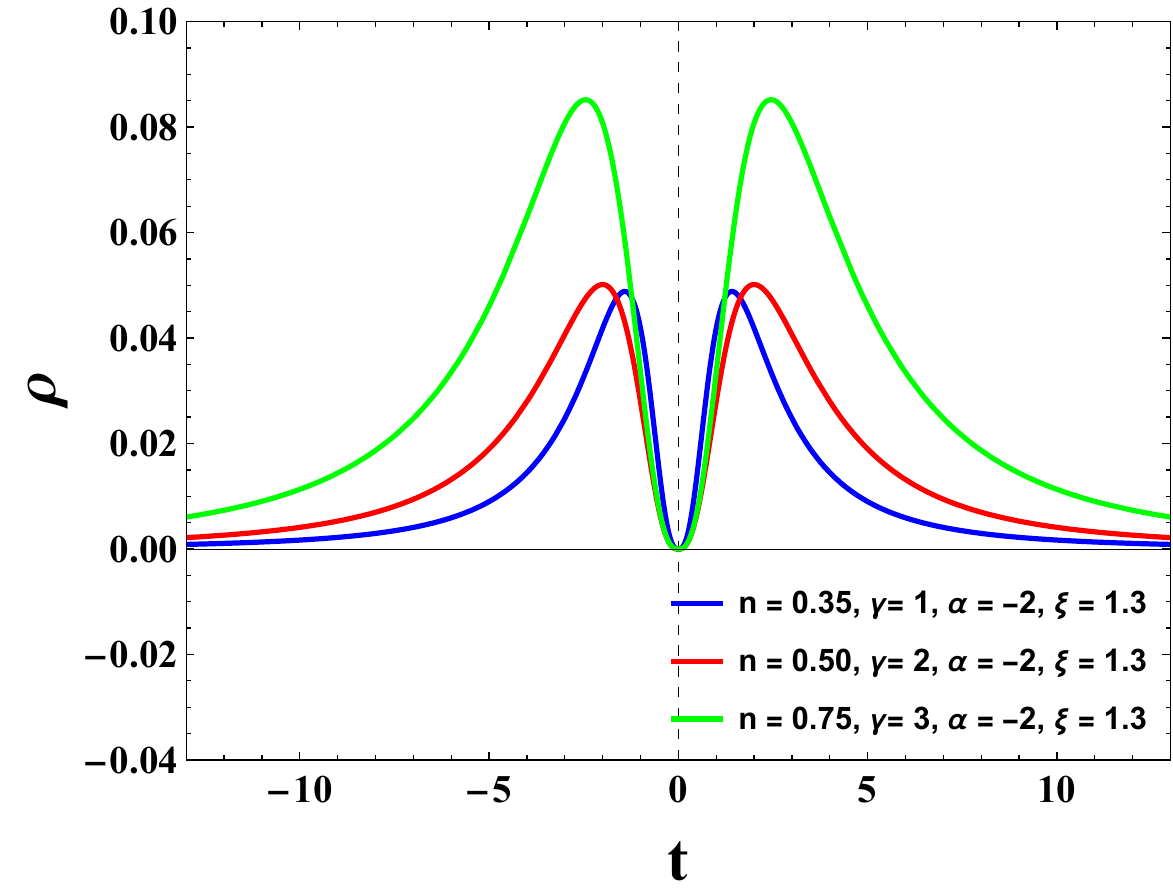}}\hfill
      \subfloat[]{\label{Eos} \includegraphics[scale=0.43]{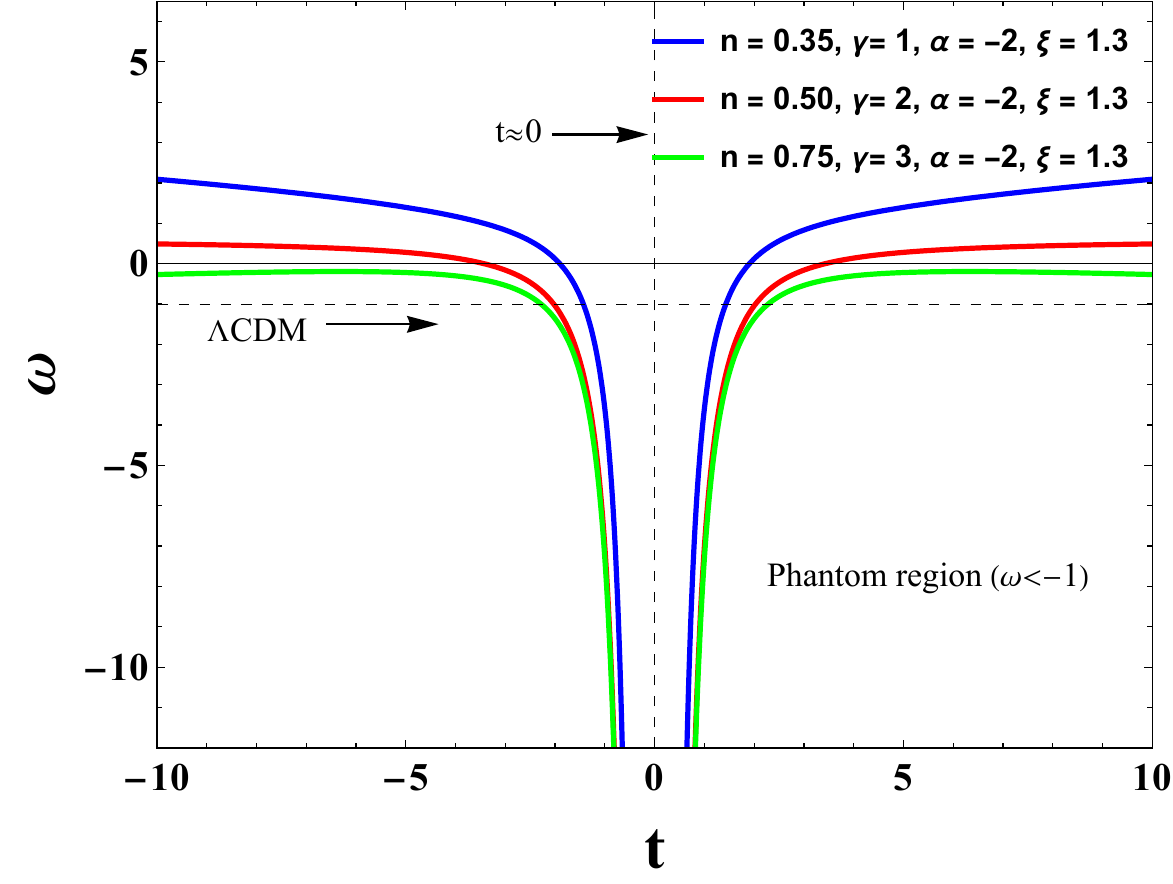}}\par
      \subfloat[]{\label{dh} \includegraphics[scale=0.43]{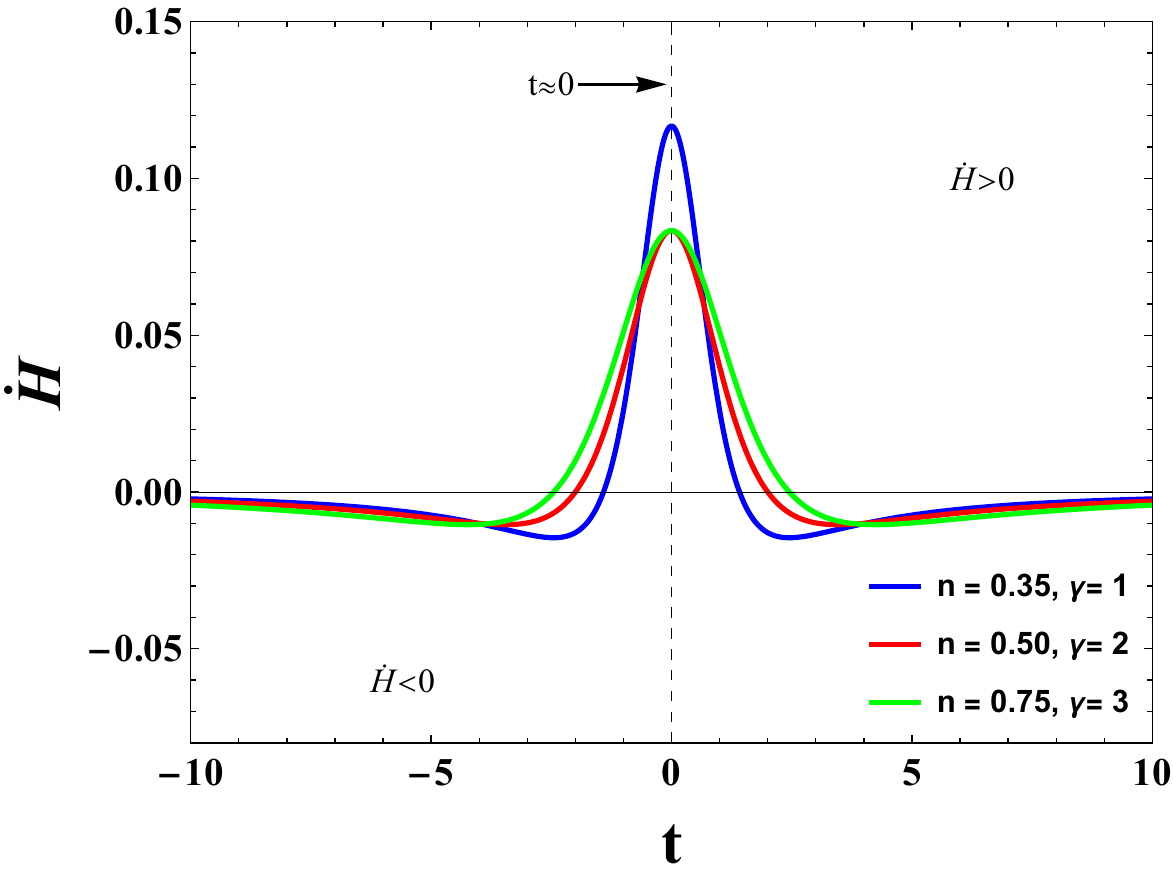}}\hfill
      \subfloat[]{\label{nec2} \includegraphics[scale=0.43]{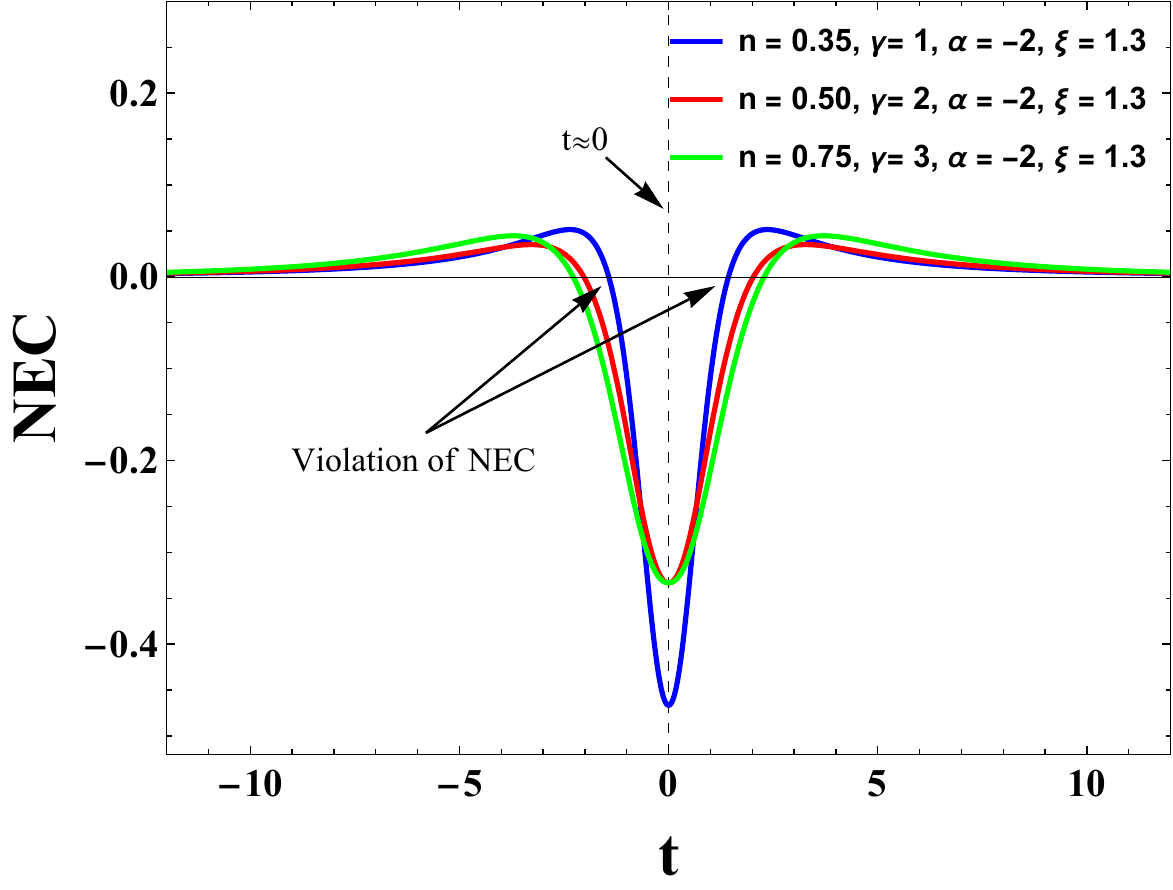}}
\end{center}
\caption{ The evolution of the dynamical status of the model from the perspective of the bouncing scenario.}
\label{fig.2}
\end{figure}

Fig.~\ref{rho(t)} reveals two key characteristics of the energy density evolution: it maintains a positive value ($\rho > 0$) throughout while asymptotically approaching zero ($\rho \to 0$) near the bouncing point at $t \approx 0$. The dynamics of the equation of state parameter $\omega$ play a pivotal role in this cosmological scenario. Our analysis shows that $\omega$ exhibits characteristic Quintom behavior, crossing the critical phantom divide ($\omega = -1$) both preceding and following the bounce, with a distinct transition into the phantom regime ($\omega < -1$) in the immediate neighborhood of the bounce point (see Fig.~\ref{Eos}). This quintessential Quintom signature confirms the model's classification as a Quintom cosmology while simultaneously highlighting the enhanced instability of the system during the bounce phase \cite{Cai:2012va}. Supporting these findings, we observe a positive Hubble parameter derivative ($\dot{H} > 0$, Fig. \ref{dh}) along with a clear violation of the null energy condition ($\rho + p < 0$, Fig.~\ref{nec2}) in the temporal vicinity of the bounce.

\subsection{ Energy Conditions}

In GR, energy conditions serve as fundamental tools for analyzing spacetime singularities and the behavior of geodesics—whether null, spacelike, timelike, or light-like. These conditions also provide critical constraints for understanding cosmic bounces, including the universe’s dynamics near the bouncing epoch and the viability of various bouncing scenarios proposed in theoretical cosmology. Derived from the Raychaudhuri equations \cite{Raychaudhuri:1953yv}, energy conditions impose limits on linear combinations of the energy density ($\rho$) and pressure ($p$). For a perfect fluid, they ensure gravity remains attractive by requiring these quantities to stay positive.

The different energy conditions, summarized in Table \ref{tabparm2}, are formulated in terms of co-oriented timelike vectors ($t^i$), spacelike vectors ($\varsigma^j$), and null (light-like) vectors ($\ell^i$). Importantly, these conditions are not independent but exhibit interrelationships \cite{Kontou:2020bta}, reflecting their shared foundation in geometric and thermodynamic principles.

In Weyl geometry, the violation of the NEC and SEC is caused by a new geometric parameter known as the non-metricity and the violation of the metricity criteria $(g_{ij;k}\ne 0)$. Unlike standard Riemannian geometry, Weyl geometry introduces a connection that causes the length of a vector to change during parallel transport $(\delta \ell =\ell_{0}w_{\alpha}\delta x^{\alpha})$. This non-metricity, or the change in length, can manifest as negative pressure, which is essential for violating the strong energy condition (SEC) and expanding behavior with cosmic acceleration.

\begin{table}[htbp]
\caption{ \textbf{Energy Conditions}}
%\begin{ruledtabular}
\begin{center}
\label{tabparm2}
\renewcommand{\arraystretch}{1.8}
\setlength{\tabcolsep}{11pt}
\begin{tabular}{lc c c r} 
\hline\hline
%\begin{tabular}{ | p{2 cm} | p{3.5cm} | p{3cm} | p{6cm} | }
% \hline
% \multicolumn{4}{|c|}{\textbf{ Energy Conditions }}

{\it Energy condition}\footnote{Null energy condition (NEC), Weak energy condition (WEC), Dominant energy condition (DEC), and Strong energy condition (SEC) \cite{Peebles:2002gy}.}   &   {\it Physical state}    &   {\it Geometric state}  &   {\it Perfect fluid state}     
 \\
 \hline
   WEC  &  $ T_{ij}t^it^j\geq 0 $  &  $  G_{ij}t^it^j\geq 0 $  &  $  \rho \geq 0, \rho+p \geq 0 $  
\\
\hline
 SEC & $ (T_{ij}-\frac{T}{n-2}g_{ij})t^it^j\geq 0  $ &  $ R_{ij}t^it^j\geq 0 $  &  $ \rho+p \geq 0, (n-3)\rho+(n-1)p \geq 0 $
\\
\hline
   DEC   &  $ T_{ij}t^i\varsigma^j\geq 0 $  &  $ G_{ij}t^i\varsigma^j\geq 0 $ &   $\rho \geq |p| $  
\\
\hline
 NEC   &  $ T_{ij}\ell^i\ell^j\geq 0 $ & $ R_{ij}\ell^i\ell^j\geq 0 $  &   $ \rho+p \geq 0 $
\\
\hline\hline
\end{tabular}    
%\end{ruledtabular} 
\end{center}
\end{table}

\begin{figure}[h]
\begin{center}
       \subfloat[]{\label{NEC} \includegraphics[scale=0.60]{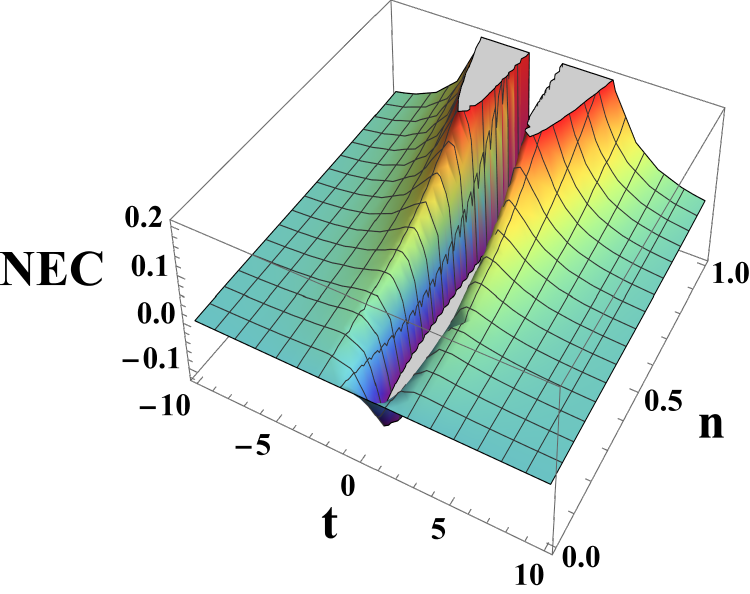}}\hfill
       \subfloat[]{\label{DEC} \includegraphics[scale=0.60]{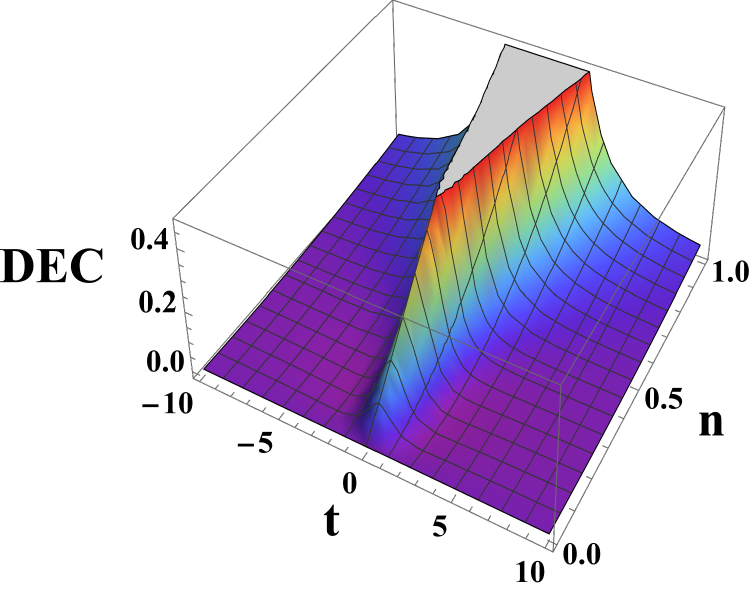}}\par
       \subfloat[]{\label{SEC} \includegraphics[scale=0.60]{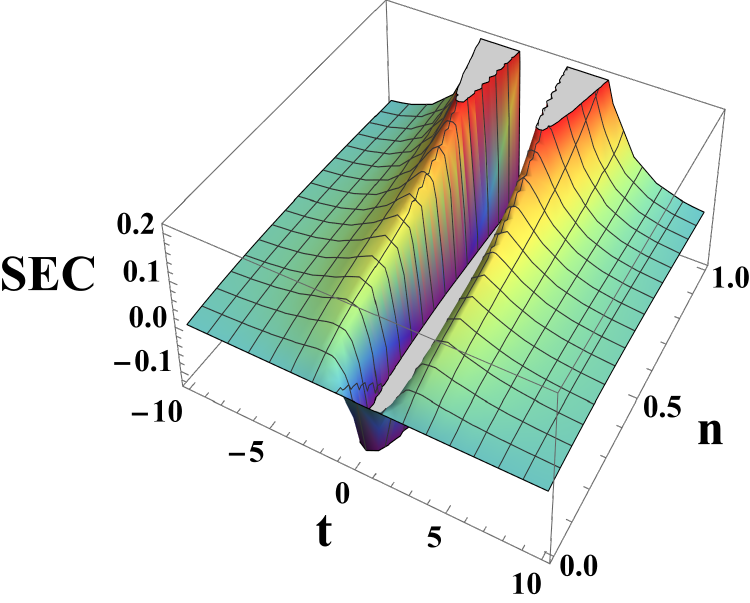}}
\end{center}
\caption{The evolution of the energy conditions in the bouncing scenario around $ t\approx0 $.}
\label{fig.3}
\end{figure}
Fig. \ref{fig.3} demonstrates the classification of energy conditions in the vicinity of the bouncing point at $ t \approx 0 $. The figures reveal an absence of singularities near the bouncing epoch, with energy conditions instead exhibiting transitional behavior around this critical region. Consistent with standard matter bounce scenarios, both the Null Energy Condition (NEC) and the Strong Energy Condition (SEC) become negative near the bounce, providing clear evidence of a violation of the energy conditions. This violation drives the model into a phantom phase characterized by an equation of state $ \omega\leq-1 $.

For a successful bouncing cosmology, two crucial requirements must be met: (1) violation of the NEC and (2) satisfaction of the condition $ \dot{H} = - 4 \pi G \rho (1+ \omega) > 0 $. Our analysis reveals that NEC is violated specifically around the bouncing point $ t \approx 0 $ for all model parameter values (Fig. \ref{NEC}), while remaining satisfied for both $ t>0 $ and $ t<0 $ \cite{Tripathy:2020atm}. Notably, the Dominant Energy Condition (DEC) holds throughout cosmic evolution (Fig. \ref{DEC}), exhibiting symmetric behavior around $t \approx 0$. This confirms the predominance of energy density in our model, ensuring its stability and physical consistency.

The violation of the SEC near $ t \approx 0 $ (Fig. \ref{SEC}) strongly suggests the presence of dark energy effects in the universe. These results collectively demonstrate that our Weyl-type $ f(Q) $ gravity model satisfies all essential requirements for a viable bouncing cosmology.

\subsection{Stability analysis}\label{V}

\qquad The stability of cosmological models in Weyl-type $f(Q)$ gravity can be investigated through the propagation speed of cosmic perturbations, analogous to sound waves in a fluid medium. This sound speed parameter characterizes how rapidly density fluctuations propagate through the cosmic medium. For a universe filled with a perfect fluid, the sound speed squared is fundamentally related to the pressure and energy density variations, defined as \cite{Peebles:2002gy}:
\begin{equation}\label{28}
C_s^2=\frac{dp}{d\rho}.
\end{equation}
A physically viable model requires thermodynamic stability, which translates to the sound speed satisfying $0 \leq C_s^2 \leq 1$, where the bounds correspond to stable perturbation propagation at sub-luminal speeds.

The speed of sound is an important factor in the linear perturbations as it decides how pressure waves propagate and influence the physical features of the system, especially in compressible fluids. Linearization simplifies the governing equations by assuming small changes, and the speed of sound is the parameter that controls the speed of the linear sound waves initially. In cosmology, the speed of sound of the dark matter or the baryonic matter affects the growth of linear perturbations. In the case of the non-zero sound speed for baryons, it causes oscillations and filtering in the power spectrum of the density fluctuations on a very small scale.

For a more rigorous stability assessment, we employ the adiabatic index $\Gamma$, a fundamental parameter introduced by Chandrasekhar \cite{Chandrasekhar:1964zz}, which governs the system's response to perturbations. This approach provides deeper insight into the dynamical stability of the model, with the adiabatic index being particularly crucial for analyzing oscillatory behavior and instability thresholds \cite{Shaily:2025kua, Chandrasekhar:1964zz, Banerjee:2024inf}.
\begin{equation}\label{28a}
    \Gamma=\big(1+\frac{\rho}{p}\big)\big(\frac{dp}{d\rho}\big)_S=\big(1+\frac{1}{\omega}\big)(\frac{dp}{d\rho}\big)_S. 
\end{equation}
In a limiting case, the stability condition provides
\[ \lim_{C_s^2\to 0} \Gamma=0 \] and \[ \lim_{C_s^2\to 1} \Gamma=1+\frac{1}{\omega}, \]
following the criterion of the speed of sound $ 0\leq C_s^2 \leq 1 $.

The speed of sound, defined as $ \frac{dp}{d\rho} $ (where the subscript $ S $ denotes constant entropy), is expressed in units of the speed of light. Crucially, the parameter $ \Gamma $ - which characterizes dynamical instabilities in relativistic systems imposes strict constraints on isotropic fluid spheres. This limiting value, known as the critical adiabatic index $ \Gamma_{\text{cr}} $, determines stability through the condition $ \langle \Gamma \rangle > \Gamma_{\text{cr}} $, where $ \langle \Gamma \rangle $ represents the averaged adiabatic index. While Newtonian gravity fixes $ \Gamma_{\text{cr}} = \frac{4}{3} $, general relativistic effects elevate $ \Gamma_{\text{cr}} $ beyond this classical threshold.

\begin{figure}\centering
\subfloat[]{\label{cs} \includegraphics[scale=0.43]{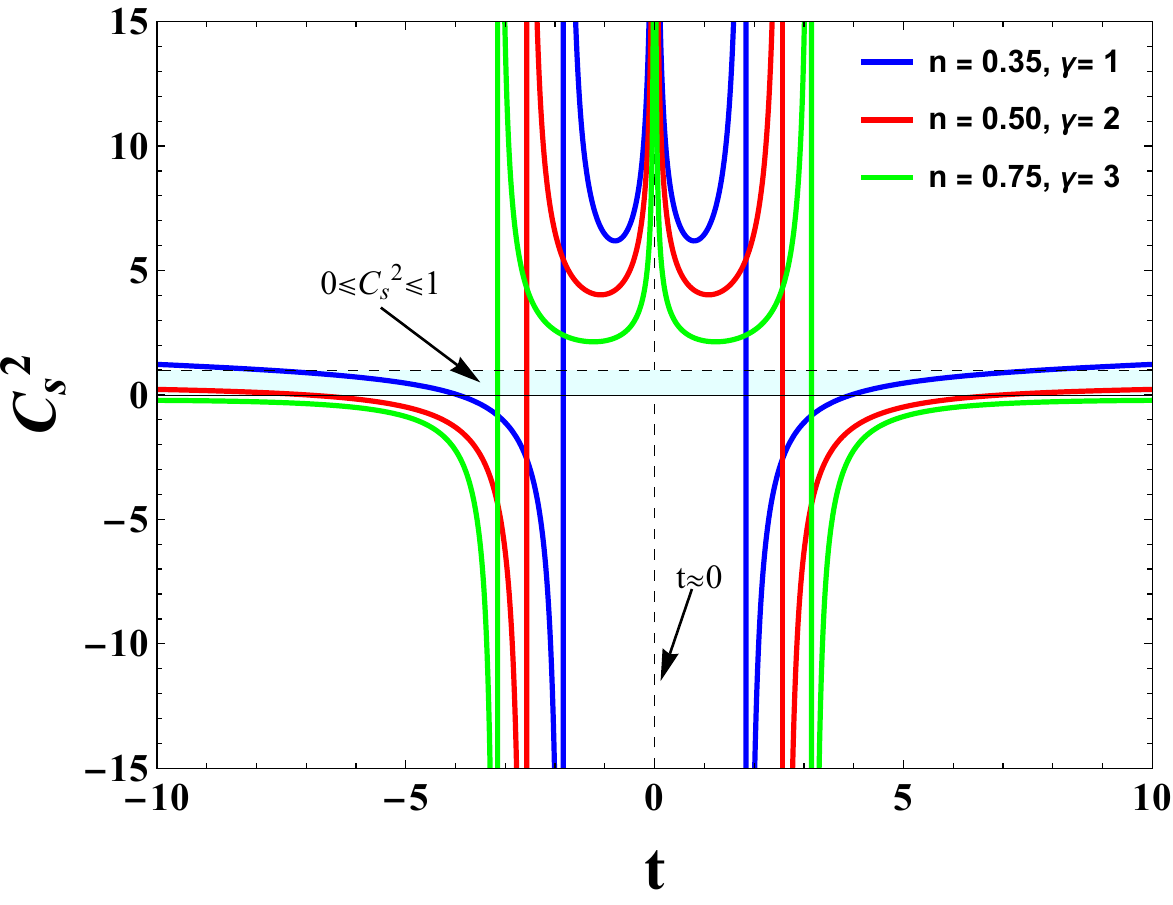}}\hfill
\subfloat[]{\label{ad} \includegraphics[scale=0.43]{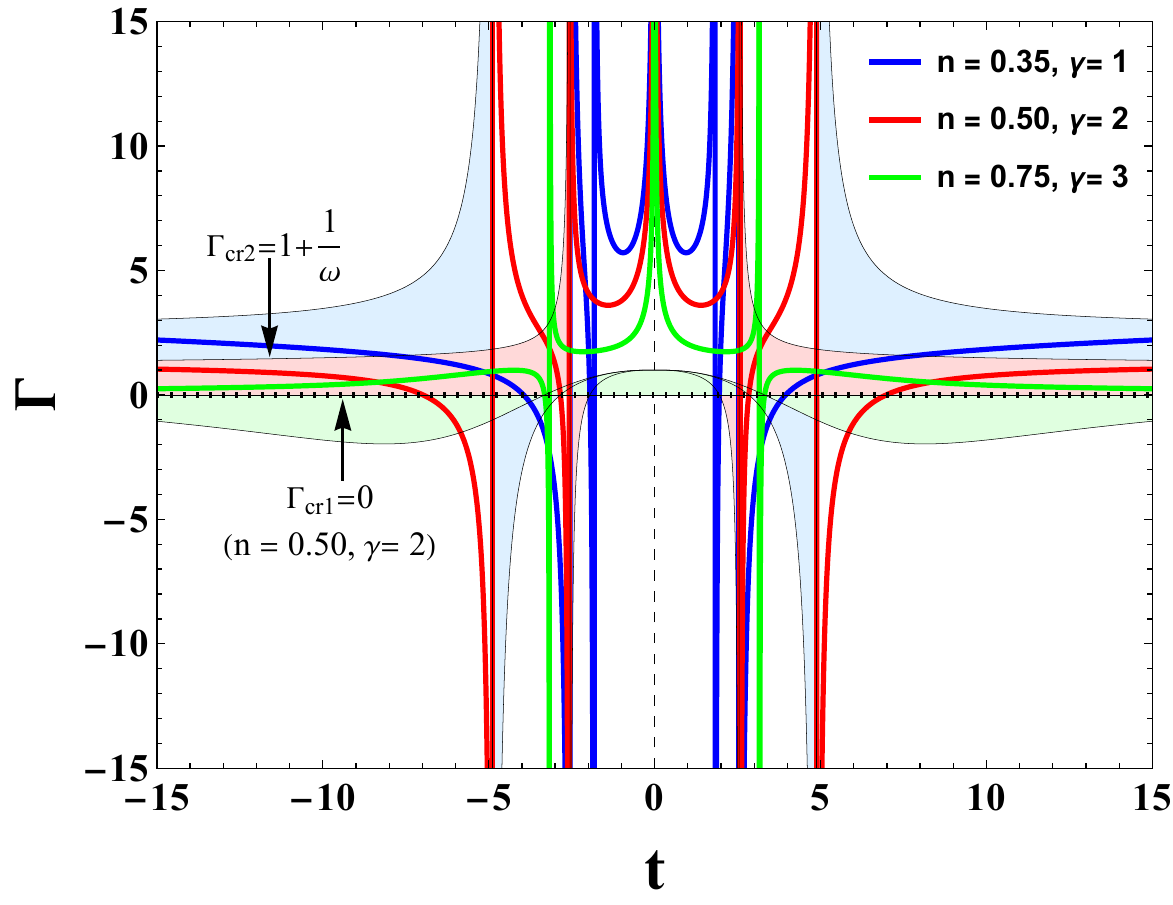}}
\caption{Evolution of stability parameters in the bouncing cosmology near $ t\approx0 $.}
\label{vos}
\end{figure}

Our analysis computes both the squared sound speed $ C_s^2 $ and adiabatic index $ \Gamma $, with their evolutionary profiles displayed in Fig. \ref{vos}. These reveal significant instability near the bouncing epoch ($ t\approx0 $). Notably, the adiabatic index maintains stability within the range $ 0\leq \Gamma \leq \left(1+ \frac{1}{\omega}\right) $ for all model parameters $ n,\gamma $.

The Gradient Instability ($ C_s^2<0)$: We acknowledge that $ C_s^2 = dp/d\rho < 0$ indicates a classical gradient instability. In our model, Fig. 4 shows that this occurs only in a very narrow interval around the bounce point ($t \approx 0$), which is a direct consequence of the NEC violation required for the bounce. The key question is whether this brief instability catastrophically amplifies perturbations. A full linear perturbation analysis is needed for a definitive answer and is proposed as future work. We rephrase our statement: “The model exhibits a short-lived gradient instability near the bounce, a common challenge in non-singular bounce models that must be carefully constrained by future perturbation studies.” 

\subsection{Effective Quintom Behavior and Scalar Field Reconstruction}\label{Va}

In GR, the quintom model has emerged as a pivotal framework for investigating bouncing cosmologies in recent years. This model uniquely bridges quintessence-like and phantom-like scalar fields by dynamically transitioning between the two regimes. While the quintom scenario with $\omega = -1$ presents observational challenges due to its analytical complexity, it becomes theoretically viable under the condition $\dot\phi^2 \ll V(\phi)$. This inequality implies that the kinetic energy (KE) of the scalar field must be negligible compared to its potential energy (PE). Notably, when $\omega \simeq -1$, this framework can describe accelerated expansion scenarios, including inflationary models, among others. We agree that introducing fundamental scalar fields on top of a modified gravity sector can be redundant. To clarify: We do not introduce new scalar fields into the action.

The scalar fields $\phi_q$ and $\phi_{ph}$ in Section \ref{Va} appear only in a reconstruction or effective description sense. The method is as follows:
\begin{enumerate}
    \item We first solve for the total effective energy density $\rho$ and pressure $p$ from the Weyl-$f(Q)$ equations.
    \item We then ask: “If one were to interpret this effective cosmic fluid as being derived from a quintom model (two scalar fields), what would the properties of those fields be?”
    \item Equations (\ref{34}-\ref{36}) are the result of this mathematical reconstruction. They show that the effective dark energy generated by the Weyl geometry and $ f(Q) $ terms mimics the behavior of a quintom model near the bounce, with an effective kinetic term that becomes negative (phantom-like) temporarily.
\end{enumerate}
This is a valuable diagnostic tool, not an extension of the theory. It helps bridge understanding between modified gravity and scalar field paradigms.

We introduce both quintessence and phantom fields to study scenarios in which dark energy crosses the phantom divide line, which means its behavior changes from quintessence-like $ \omega> -1$ to phantom-like $ \omega<-1 $ at the bounce. While modified gravity can produce phantom-like behavior, it often does so with inherent instabilities and is generally more stable when a single field can transition across the phantom divide. Studying quintom models with both fields allows researchers to build more realistic and stable models for the universe's accelerated expansion. We consider different field types that allow for more complex and comprehensive models that can better capture the nuances of dark energy and the bouncing scenario in the modified gravity. This helps to explore a wider range of possibilities for the evolution of the universe and better constrain the underlying physics. The explicit scalar fields are needed to support the successful bouncing scenario in the modified gravity.  

The modified Weyl $f(Q)$ gravity, while providing new geometric degrees of freedom in the Weyl connection, often struggles to perfectly mimic $\Lambda$CDM on its own due to potential instabilities, doubling degrees of freedom by introducing an extra scalar field like in $ f(Q,T) $ gravity or coupled models or promoting the Weyl field dynamically provides more parameters and flexibility to fit diverse cosmological data like dark energy or bounce dynamics while remaining ghost-free and consistent with second-order field equations, justifying the extra complexity to solve problems general relativity can not, say the Hubble tension.

The link between scalar fields and Weyl geometry is geometrical, where the scalar field $\phi$ determines the length scale in a Weyl integrable spacetime, defining a non-metricity condition on the metric tensor $g_{ij }$. This condition, derived from the Palatini variation of an action, implies that the connection is not compatible with the metric in the standard Riemannian sense, i.e., the length of a vector changes under parallel transport. The scalar field's role is to modify the geometry of spacetime, which in turn affects the dynamics of the fields, connecting gravity, electromagnetism, and matter within a single geometric framework.

The gravitational action in Einstein's theory is expressed as
\begin{equation}{\label{29}}
\mathcal{A} = \frac{c^4}{16 \pi G} \int{R \sqrt{-g} d^4 x + \mathcal{A}m},
\end{equation}
where $\mathcal{A}m$ incorporates the actions for both quintessence-like ($\mathcal{A}_{q}$) and phantom-like ($\mathcal{A}_{ph}$) scalar fields. Here, the speed of light $c$ is normalized to unity ($c=1$) for simplicity.
The actions for quintessence-like and phantom-like scalar fields are given by 
\begin{align}
   \mathcal{A}_{q} =& \int \bigg( -\frac{1}{2} \partial_i \phi_{q} \partial^i \phi _{q} -V(\phi_{q}) \bigg) \sqrt{-g} d^4 x, \label{30}\\ 
   \mathcal{A}_{ph} =& \int \bigg( \frac{1}{2} \partial_i \phi_{ph} \partial^i \phi _{ph} -V(\phi_{ph}) \bigg) \sqrt{-g} d^4 x \label{31}
\end{align}
respectively. 

Therefore, the energy density and total pressure of the Universe filled with quintessence-like and phantom-like scalar fields are obtained as \cite{Singh:2018xjv, Gibbons:2003yj}
\begin{align}
     \rho_{q} = \frac{1}{2} \dot{{\phi}^2}_{q} + V(\phi_{q}), \hskip0.5in p_{q} = \frac{1}{2} \dot{{\phi}^2}_{q} - V(\phi_{q}) \label{32}.\\
       \rho_{ph} = -\frac{1}{2} \dot{{\phi}^2}_{ph} + V(\phi_{ph}), \hskip0.5in p_{ph} = -\frac{1}{2} \dot{{\phi}^2}_{ph} - V(\phi_{ph}) \label{33}.
\end{align}

From equations (\ref{22}) and (\ref{23}), the kinetic and potential energies of the quintessence-like scalar field and the phantom-like scalar field are evaluated as \cite{Barrow:1988xh}
\begin{align}\label{34}
    \frac{1}{2}\dot{\phi^2_{q}} =& -\alpha \xi  \left(\frac{8}{3}\right)^{\xi} \left(\frac{\beta ^2 n^2 t^2}{\left(\gamma +\beta  t^2\right)^2}\right)^{\xi} +\frac{2 \beta  n \left(\beta  t^2 \left(-\left(\left(m^2+12\right) n-6\right)\right)-6 \gamma \right)}{9 \left(\gamma +\beta  t^2\right)^2}, \end{align}
\begin{align}\label{35}    
    \frac{1}{2}\dot{\phi^2_{ph}} =& \alpha \xi  \left(\frac{8}{3}\right)^{\xi} \left(\frac{\beta ^2 n^2 t^2}{\left(\gamma +\beta  t^2\right)^2}\right)^{\xi} +\frac{2 \beta  n \left(\beta  t^2 \left(-\left(\left(m^2+12\right) n-6\right)\right)-6 \gamma \right)}{9 \left(\gamma +\beta  t^2\right)^2}, \end{align}
 \begin{align}\label{36}
  V(\phi_{q})=V(\phi_{ph}) =& -\alpha \xi \bigg(\frac{8}{3}\bigg)^{\xi}  \left(\frac{\beta ^2 n^2 t^2}{\left(\gamma +\beta  t^2\right)^2}\right)^{\xi} + \frac{4 \beta  n \left(\gamma +\beta  (2 n-1) t^2\right)}{3 \left(\gamma +\beta  t^2\right)^2} .
\end{align}

The EoS parameter $ \omega $ for the quintessence-like and phantom-like scalar fields is given by
\begin{equation}{\label{37}}
    \omega = \frac{p_{\phi_{q}}}{\rho_{\phi_{q}}} >-1, \hskip0.2in \omega = \frac{p_{\phi_{ph}}}{\rho_{\phi_{ph}}} <-1.
\end{equation}

\begin{figure}
\begin{center}
       \subfloat[]{\label{phiqu} \includegraphics[scale=0.4]{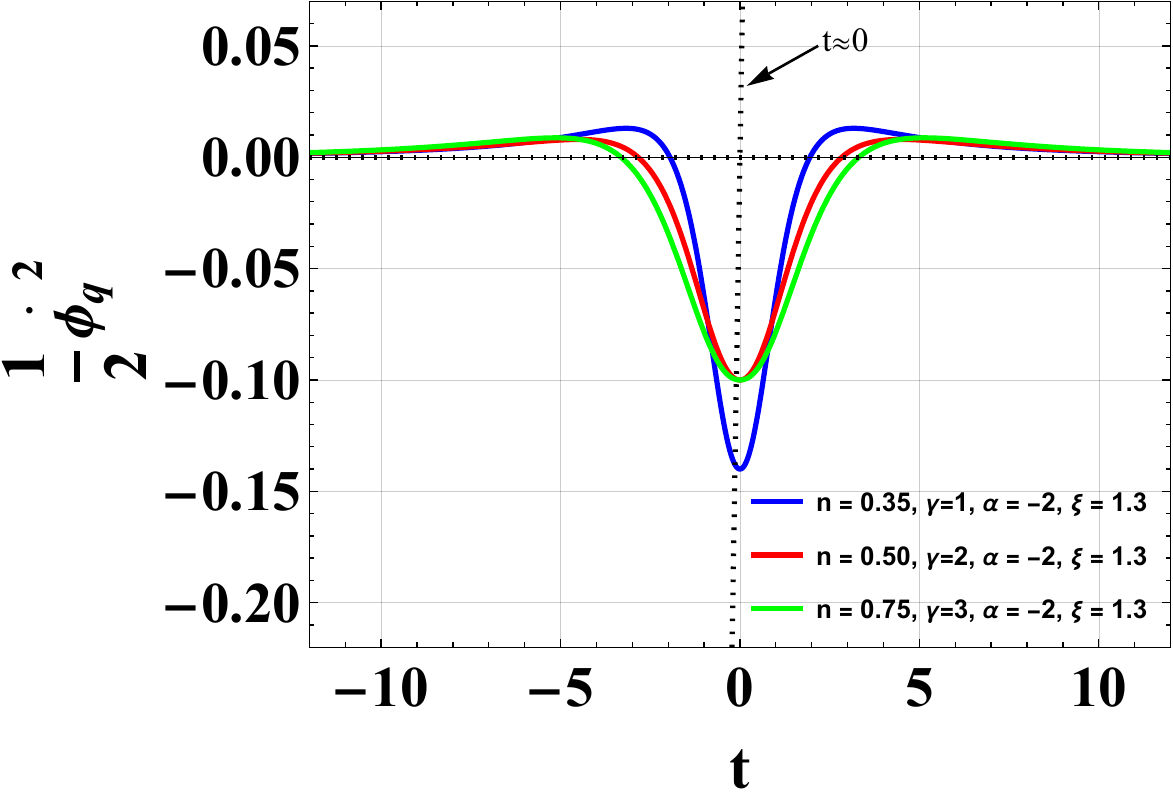}}\hfill
       \subfloat[]{\label{phiph} \includegraphics[scale=0.4]{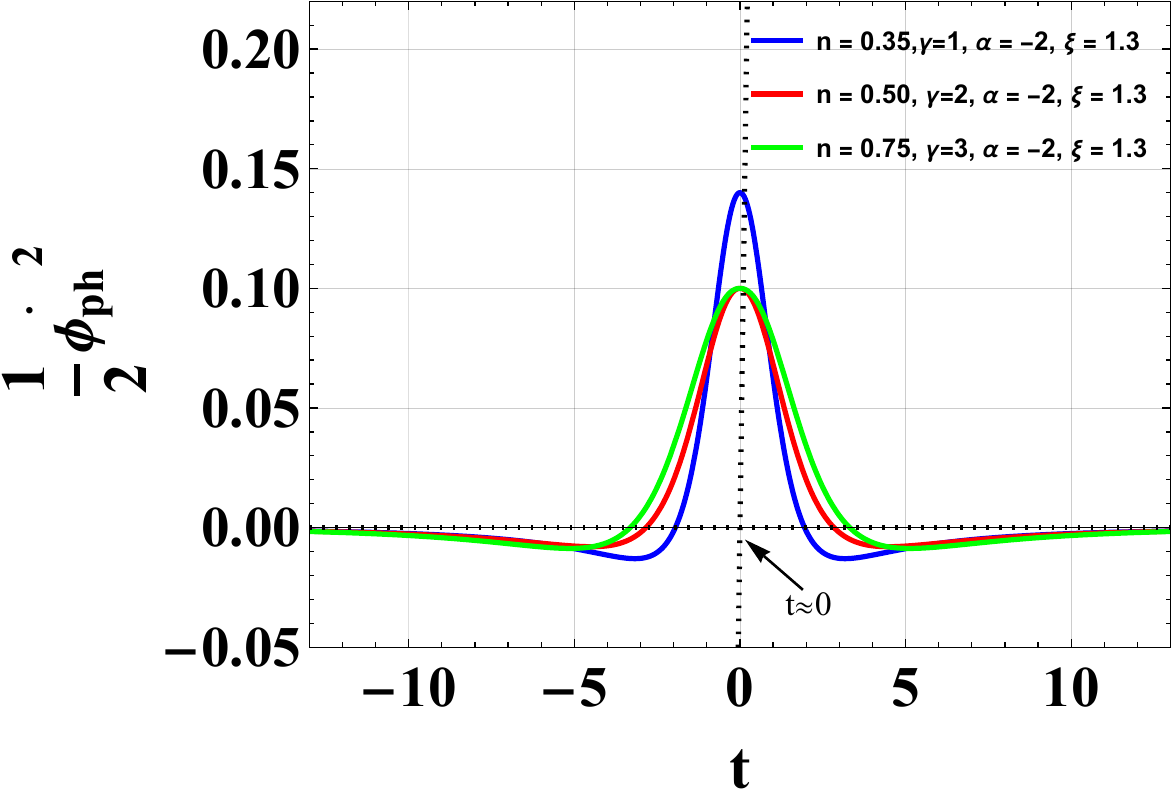}}\par
     \subfloat[]{\label{Vphi} \includegraphics[scale=0.43]{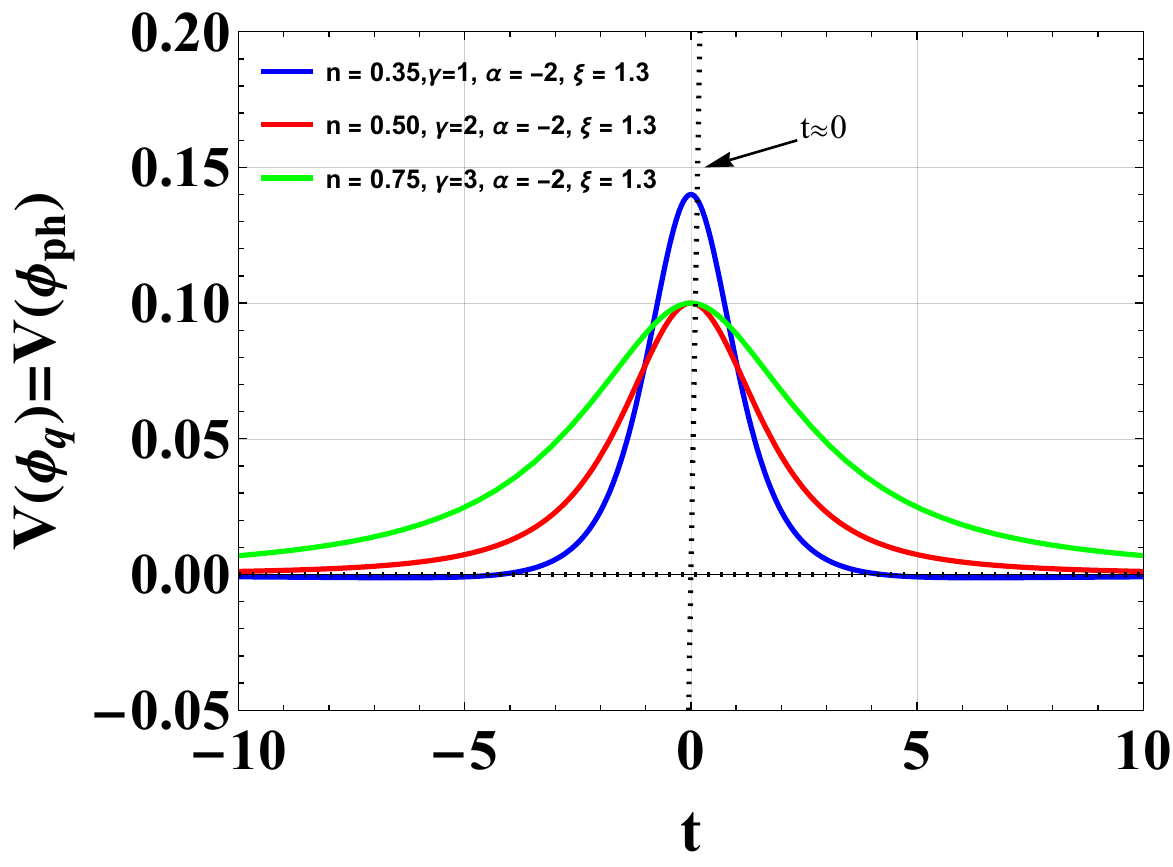}}
\end{center}
\caption{The evolution of Scalar Fields in the bouncing scenario around $ t=0 $ for $ \beta =0.3 $, and $ m=0.6 $.}
\label{sf}
\end{figure}

Equations (\ref{32}) and (\ref{33}) yield the kinetic energy expressions for the quintessence-like and phantom-like scalar fields as $\frac{1}{2} \dot{\phi^2_{q}} = \rho_{q} + p_{q}$ and $\frac{1}{2} \dot{\phi^2_{ph}} = -(\rho_{ph} + p_{ph})$, respectively. Furthermore, analysis of Eqs. (\ref{34}-\ref{36}) reveals a crucial feature: when the equation-of-state parameter $\omega$ crosses the phantom divide line ($\omega = -1$), the kinetic energies of both scalar fields become numerically equivalent. This equality represents a fundamental requirement for realizing the bouncing scenario in our model \cite{Cai:2007qw, Cai:2008ed, Shabani:2017lgy}.

The dark energy and dark matter components can be effectively modeled through a scalar field framework incorporating negative kinetic energy. As is evident from Eq. (\ref{33}), the phantom-like kinetic energy takes the form $(\frac{1}{2}\dot{\phi_{ph}}^2)=-(\rho_{\phi_{ph}}+p_{\phi_{ph}})$. Such phantom field models with negative kinetic energy have been extensively studied in cosmological contexts \cite{Gibbons:2003yj, Nojiri:2003ag, Caldwell:1999ew, Cline:2003gs}, particularly in relation to cosmic acceleration.

Our investigation extends to the comparative analysis of quintessence-like kinetic energy ($\frac{1}{2} \dot{\phi^2_{q}}$), phantom-like kinetic energy $ (\frac{1}{2}\dot{\phi_{ph}}^2) $, and their respective potential energies $V(\phi_q)$ and $V(\phi_{ph})$ as functions of cosmic time $ t $ within the Weyl-type $f(Q)$ gravity framework. These relationships are graphically represented in Fig. \ref{sf} for various model parameters.

In Weyl-type $f(Q)$ gravity, the behavior of kinetic energy near the bouncing point ($t \approx 0$) exhibits distinctive characteristics for different scalar fields \cite{Cai:2012va}. The quintessence-like scalar field attains its minimum negative kinetic energy value, while the phantom-like scalar field reaches its maximum positive kinetic energy in this regime \cite{Bamba:2008ut}. The negative kinetic energy of the quintessence-like field manifests itself as a dark energy effect, producing repulsive forces near the bounce point $t \approx 0$ (see Figs. \ref{phiqu}, \ref{phiph}), consistent with the quintom behavior observed in bouncing cosmologies \cite{Cai:2008ed}.

Analysis of Fig. \ref{Vphi} reveals that both the quintessence potential $V(\phi_q)$ and the phantom potential $V(\phi_{ph})$ achieve extremum values in the vicinity of the bouncing point at $ t \approx 0 $ within this Weyl-type $f(Q)$ gravity framework. This extreme behaviour of potentials correlates with the characteristic energy condition violations required for successful bounce scenarios \cite{Novello:2008ra}.

A model with a negative kinetic term, which is the characteristic of a ghost field, is considered unstable in quantum field theory due to the potential for catastrophic vacuum decay and can not be considered physically acceptable in general. In this model, the negativity indicates instability near the bounce as it changes its state from a contracting to an expanding state of the Universe. Therefore, it may be considered as physically viable in the case of the bouncing Scenario.

Negative Kinetic Energy and Ghosts: The negative value of $ \frac{1}{2} \dot{\phi_q}^2$ obtained in the effective scalar field reconstruction does not imply a fundamental ghost degree of freedom in our theory. Ghosts are defined at the level of the fundamental action. Our action Eq. (\ref{1}) for the metric and Weyl vector is ghost-free \cite{Barvinsky:2013mea, BeltranJimenez:2016wxw, Lin:2020phk}. The negative effective kinetic energy is a phenomenological artifact of the reconstruction, indicating that the total effective fluid behaves like a phantom field near the bounce. This is a feature of the effective description, not of the fundamental Lagrangian.

The reconstructed scalar fields are an effective description only. The negative kinetic energy does not indicate a ghost instability in the underlying Weyl-type $f(Q)$ theory, which is known to be ghost-free.

\section{ Concluding Remarks}{\label{sec-4}}
This work establishes Weyl-type $ f(Q) $ gravity as a viable framework for singularity-free bouncing cosmologies. The model satisfies all theoretical requirements for a successful bounce:
\begin{enumerate}
    \item Nonsingular scale factor evolution with a smooth contraction-to-expansion transition.
    \item NEC violation and $ \dot{H} > 0 $ near the bounce.
    \item Quintom-like EoS behavior crosses $ \omega = -1 $.
    \item Finite energy density and curvature invariants throughout the bounce.
\end{enumerate}
The analysis of scalar fields further validates the model's alignment with dark energy dynamics, while stability studies highlight transient instabilities inherent to bounce transitions. Future work will explore perturbations, observational signatures (e.g., primordial gravitational waves), and constraints from cosmological data. This model enriches the landscape of modified gravity theories and provides a compelling alternative to inflationary paradigms to resolve the initial singularity problem.

This work extensively examines the bouncing cosmological model of the Universe in the Weyl-type $ f(Q) $ gravity with the non-metricity tensor $ Q $ within the framework of the flat FLRW model. We have considered a simple power functional form of $ f(Q) =  \alpha Q^{\xi} $, where $ \alpha $ and $ \xi $ are model parameters. Furthermore, to discuss the specific behavior of the bouncing cosmology, we have taken an ansatz of scale factor correlated with the bounce \cite{Odintsov:2016tar}. Our model experiences a bounce at $ t\approx0 $ based on the examination of the various dynamical parameters. The variation of the scale factor $ a $ attains the non-vanishing minimum value at the bouncing position $ t\approx0 $ as observed in Fig. \ref{a(t)}.

The Hubble parameter $ H $ is an essential cosmic parameter in analyzing the model's bouncing behavior. In Fig. \ref{H(t)}, it is noticeable that $ H $ undergoes a phase shift from a contracting state to an expansive regime at the locality of the bouncing point $ t\approx0 $. This transition shows contraction when $ H(t) <0 $ and expansion when $ H(t) >0 $. 
 
The deceleration parameter $q $ transitions from deceleration to acceleration before the bounce and from acceleration to deceleration after the bounce. This model shows the inflationary state of the universe in a very short interval of time after the bounce (see Fig. \ref{q(t)}). The variation of Hubble radius $ r_h $ can also be seen in the neighborhood of the bouncing point $ t\approx0 $ in Fig. \ref{Hr(t)}. 

We have studied the dynamical parameters, including the energy density $ \rho $, isotropic pressure $ p $, and EoS in the bouncing scenario for the specific model. The energy density $ \rho>0 $ during the evolution and $ \rho \to0 $ at the bouncing point (see Fig. \ref{rho(t)}). Fig. \ref{Eos} exhibits that the model is a Quintom model as the EoS parameter $ \omega(t) $ crosses the phantom divide line $ \omega = -1 $. It also undergoes a phase transition from a perfect fluid state to a dark energy state and vice versa in the vicinity of the bouncing point $ t\approx0 $. This model shows that $ \dot{H}>0 $ and violates the NEC near the bouncing point $ t\approx0 $, which satisfies the essential features of the bouncing model (see Figs. \ref{dh} and \ref{nec2}). Additionally, Fig. \ref{fig.3} demonstrates that the DEC is satisfied, while NEC and SEC do not satisfy the neighbourhood of the bouncing point $ t\approx0 $. A consequence of the brief NEC violation could be a suppression of the CMB power spectrum on the largest scales or specific non-Gaussian signatures, which could be tested against future data from observatories like LiteBIRD. Figure \ref{vos} exhibits that the model is highly unstable near the bounce according to the view of the speed of sound and the adiabatic index.
 
The kinetic energy of quintessence-like scalar fields is minimum negative, whereas the kinetic energy of phantom-like scalar fields is maximum positive near the bouncing point.  The negative values of the quintessence-like kinetic energy of the scalar field show the dark energy model due to the repulsive force near the bounce. The potential energies $ V(\phi_q) $ and $ V(\phi_{ph}) $ for quintessence and phantom-like scalar field are extremum in the neighborhood of the bouncing point at $ t \approx 0 $. Finally, we find that our model is a non-singular bouncing model and behaves like a quintom model in the Weyl-type $ f(Q) $ gravity.

\section*{Acknowledgement} The authors express their sincere gratitude to the respected referees for their valuable comments and suggestions. The authors thank the Department of Mathematics, NSUT, New Delhi, India, for providing the necessary facilities where a part of this paper was completed. 
%The authors express their sincere gratitude to the respected referees for their valuable comments and suggestions. 

\section*{Data Availability Statement} In this manuscript, we have neither used any observational data nor produced any form of new data.

\section*{Conflicts of Interest} The authors assert that there are no conflicts of interest in the publication of this work.


\begin{thebibliography}{99}

%\cite{SupernovaSearchTeam:1998fmf}
\bibitem{SupernovaSearchTeam:1998fmf}
A.~G.~Riess \textit{et al.} [Supernova Search Team],
%``Observational evidence from supernovae for an accelerating universe and a cosmological constant,''
Astron. J. \textbf{116}, 1009-1038 (1998).
%doi:10.1086/300499
%[arXiv:astro-ph/9805201 [astro-ph]].

\bibitem{perlmutter2003measuring}
Perlmutter, Saul and Schmidt, Brian P,
% Measuring cosmology with supernovae
Supernovae and Gamma-Ray Bursters, \textbf{195}, 217 (2003).

%\cite{WMAP:2003elm}
\bibitem{WMAP:2003elm}
D.~N.~Spergel \textit{et al.} [WMAP],
%``First year Wilkinson Microwave Anisotropy Probe (WMAP) observations: Determination of cosmological parameters,''
Astrophys. J. Suppl. \textbf{148}, 175-194 (2003).
%doi:10.1086/377226
%[arXiv:astro-ph/0302209 [astro-ph]].

%\cite{WMAP:2006bqn}
\bibitem{WMAP:2006bqn}
D.~N.~Spergel \textit{et al.} [WMAP],
%``Wilkinson Microwave Anisotropy Probe (WMAP) three year results: implications for cosmology,''
Astrophys. J. Suppl. \textbf{170}, 377 (2007).
%doi:10.1086/513700
%[arXiv:astro-ph/0603449 [astro-ph]].

%\cite{WMAP:2003ogi}
\bibitem{WMAP:2003ogi}
C.~L.~Bennett \textit{et al.} [WMAP],
%``The Microwave Anisotropy Probe (MAP) mission,''
Astrophys. J. \textbf{583}, 1-23 (2003).
%doi:10.1086/345346
%[arXiv:astro-ph/0301158 [astro-ph]].


%\cite{BICEP2:2014owc}
\bibitem{BICEP2:2014owc}
P.~A.~R.~Ade \textit{et al.} [BICEP2],
%``Detection of $B$-Mode Polarization at Degree Angular Scales by BICEP2,''
Phys. Rev. Lett. \textbf{112}, no.24, 241101 (2014).
%doi:10.1103/PhysRevLett.112.241101
%[arXiv:1403.3985 [astro-ph.CO]].

%%%%%%%% for Q
\bibitem{Clifton:2011jh}
T.~Clifton, P.~G.~Ferreira, A.~Padilla and C.~Skordis,
%``Modified Gravity and Cosmology,''
Phys. Rept. \textbf{513}, 1-189 (2012).

%\cite{Peebles:2002gy}
\bibitem{Peebles:2002gy}
P.~J.~E.~Peebles and B.~Ratra,
%``The Cosmological Constant and Dark Energy,''
Rev. Mod. Phys. \textbf{75}, 559-606 (2003).
%doi:10.1103/RevModPhys.75.559
%[arXiv:astro-ph/0207347 [astro-ph]].

\bibitem{Sahni:1999gb}
V.~Sahni and A.~A.~Starobinsky,
%``The Case for a positive cosmological Lambda term,''
Int. J. Mod. Phys. D \textbf{9}, 373-444 (2000).

%\cite{BeltranJimenez:2017tkd}
\bibitem{BeltranJimenez:2017tkd}
J.~Beltr\'an Jim\'enez, L.~Heisenberg and T.~Koivisto,
%``Coincident General Relativity,''
Phys. Rev. D \textbf{98}, no.4, 044048 (2018).
%doi:10.1103/PhysRevD.98.044048
%[arXiv:1710.03116 [gr-qc]].

%% f(R) 
%\cite{Starobinsky:2007hu}
\bibitem{Starobinsky:2007hu}
A.~A.~Starobinsky,
%``Disappearing cosmological constant in f(R) gravity,''
JETP Lett. \textbf{86}, 157-163 (2007).
%doi:10.1134/S0021364007150027
%[arXiv:0706.2041 [astro-ph]].

%\cite{Capozziello:2008qc}
\bibitem{Capozziello:2008qc}
S.~Capozziello, V.~F.~Cardone and V.~Salzano,
%``Cosmography of f(R) gravity,''
Phys. Rev. D \textbf{78}, 063504 (2008).
%doi:10.1103/PhysRevD.78.063504
%[arXiv:0802.1583 [astro-ph]].

\bibitem{Goswami:2022vfq}
G.~K.~Goswami, R.~Rani, H.~Balhara and J.~K.~Singh,
``Curvature dominance dark energy model in f(R)-gravity,''
Indian J. Phys. \textbf{97}, no.12, 3707-3714 (2023)
%doi:10.1007/s12648-023-02674-3.


%\cite{Wu:2015maa}
\bibitem{Wu:2015maa}
B.~Wu and B.~Q.~Ma,
%``Spherically symmetric solution of $f(R,\mathcal{G})$ gravity at low energy,''
Phys. Rev. D \textbf{92}, no.4, 044012 (2015).
%doi:10.1103/PhysRevD.92.044012
%[arXiv:1510.08552 [gr-qc]].

%\cite{Naz:2023pfl}
\bibitem{Naz:2023pfl}
T.~Naz, A.~Malik, M.~K.~Asif and I.~Fayyaz,
%``Evolving embedded traversable wormholes in f(R,G)~gravity: A comparative study,''
Phys. Dark Univ. \textbf{42}, 101301 (2023).
%doi:10.1016/j.dark.2023.101301

\bibitem{Rani:2024uah}
R.~Rani, Shaily, G.~K.~Goswami and J.~K.~Singh,
%``Power law cosmology in Gauss-Bonnet gravity with pragmatic analysis,''
JHEAp \textbf{45}, 168-180 (2025).

\bibitem{Harko:2011kv}
T.~Harko, F.~S.~N.~Lobo, S.~Nojiri and S.~D.~Odintsov,
%``$f(R,T)$ gravity,''
Phys. Rev. D \textbf{84}, 024020 (2011).

\bibitem{Singh:2022eun}
J.~K.~Singh, A.~Singh, G.~K.~Goswami and J.~Jena,
%``Dynamics of a parametrized dark energy model in f(R,T) gravity,''
Annals Phys. \textbf{443}, 168958 (2022).

%\cite{Singh:2024ckh}
\bibitem{Singh:2024ckh}
J.~K.~Singh, Shaily, H.~Balhara, S.~G.~Ghosh and S.~D.~Maharaj,
%``EDSFD parameterization in f(R,T) gravity with linear curvature terms,''
Phys. Dark Univ. \textbf{45} (2024), 101513.
%doi:10.1016/j.dark.2024.101513
%0 citations counted in INSPIRE as of 16 May 2024

%\cite{Singh:2024kez}
\bibitem{Singh:2024kez}
J.~K.~Singh, H.~Balhara, Shaily and P.~Singh,
%``The constrained accelerating universe in f(R,T) gravity,''
Astron. Comput. \textbf{46} (2024), 100795.
%doi:10.1016/j.ascom.2024.100795
%26 citations counted in INSPIRE as of 06 Dec 2025

%\cite{Capozziello:2011hj}
\bibitem{Capozziello:2011hj}
S.~Capozziello, V.~F.~Cardone, H.~Farajollahi and A.~Ravanpak,
%``Cosmography in f(T)-gravity,''
Phys. Rev. D \textbf{84}, 043527 (2011).
%doi:10.1103/PhysRevD.84.043527
%[arXiv:1108.2789 [astro-ph.CO]].

%\cite{Cai:2015emx}
\bibitem{Cai:2015emx}
Y.~F.~Cai, S.~Capozziello, M.~De Laurentis and E.~N.~Saridakis,
%``f(T) teleparallel gravity and cosmology,''
Rept. Prog. Phys. \textbf{79}, no.10, 106901 (2016).
%doi:10.1088/0034-4885/79/10/106901
%[arXiv:1511.07586 [gr-qc]].

\bibitem{Bahamonde:2015zma}
S.~Bahamonde, C.~G.~B{\"o}hmer and M.~Wright,
%``Modified teleparallel theories of gravity,''
Phys. Rev. D \textbf{92}, no.10, 104042 (2015)
%doi:10.1103/PhysRevD.92.104042

%\cite{Heisenberg:2023lru}
\bibitem{Heisenberg:2023lru}
L.~Heisenberg,
%``Review on f(Q) gravity,''
Phys. Rept. \textbf{1066}, 1-78 (2024).
%doi:10.1016/j.physrep.2024.02.001
%[arXiv:2309.15958 [gr-qc]].

\bibitem{Lin:2021uqa}
R.~H.~Lin and X.~H.~Zhai,
%``Spherically symmetric configuration in $f(Q)$ gravity,''
Phys. Rev. D \textbf{103}, no.12, 124001 (2021).

\bibitem{Lazkoz:2019sjl}
R.~Lazkoz, F.~S.~N.~Lobo, M.~Ortiz-Ba\~nos and V.~Salzano,
%``Observational constraints of $f(Q)$ gravity,''
Phys. Rev. D \textbf{100}, no.10, 104027 (2019).

%\cite{Frusciante:2021sio}
\bibitem{Frusciante:2021sio}
N.~Frusciante,
%``Signatures of $f(Q)$-gravity in cosmology,''
Phys. Rev. D \textbf{103}, no.4, 044021 (2021).
%doi:10.1103/PhysRevD.103.044021
%[arXiv:2101.09242 [astro-ph.CO]].

%\cite{Mandal:2020lyq}
\bibitem{Mandal:2020lyq}
S.~Mandal, P.~K.~Sahoo and J.~R.~L.~Santos,
%``Energy conditions in $f(Q)$ gravity,''
Phys. Rev. D \textbf{102}, no.2, 024057 (2020).
%doi:10.1103/PhysRevD.102.024057
%[arXiv:2008.01563 [gr-qc]].

\bibitem{DAmbrosio:2021zpm}
F.~D'Ambrosio, S.~D.~B.~Fell, L.~Heisenberg and S.~Kuhn,
%``Black holes in f(Q) gravity,''
Phys. Rev. D \textbf{105}, no.2, 024042 (2022).

\bibitem{Sokoliuk:2023pby}
O.~Sokoliuk, S.~Pradhan, A.~Baransky and P.~K.~Sahoo,
%``AdS Black Hole Thermodynamics and Microstructures from f(Q) Gravitation,''
Fortsch. Phys. \textbf{72}, no.1, 2300043 (2024)
%doi:10.1002/prop.202300043

\bibitem{Bajardi:2023vcc}
F.~Bajardi and S.~Capozziello,
%``Minisuperspace quantum cosmology in f(Q) gravity,''
Eur. Phys. J. C \textbf{83}, no.6, 531 (2023)
%doi:10.1140/epjc/s10052-023-11703-8

\bibitem{Haghani:2012bt}
Z.~Haghani, T.~Harko, H.~R.~Sepangi and S.~Shahidi,
%``Weyl-Cartan-Weitzenboeck gravity as a generalization of teleparallel gravity,''
JCAP \textbf{10}, 061 (2012).

\bibitem{Xu:2020yeg}
Y.~Xu, T.~Harko, S.~Shahidi and S.~D.~Liang,
%``Weyl type $f(Q,T)$ gravity, and its cosmological implications,''
Eur. Phys. J. C \textbf{80}, no.5, 449 (2020).

%\cite{Goswami:2023knh}
\bibitem{Goswami:2023knh}
G.~K.~Goswami, R.~Rani, J.~K.~Singh and A.~Pradhan,
%``FLRW cosmology in Weyl type f(Q) gravity and observational constraints,''
JHEAp \textbf{43}, 105-113 (2024).
%doi:10.1016/j.jheap.2024.06.011
%[arXiv:2309.01233 [gr-qc]].

\bibitem{Guth:1980zm}
A.~H.~Guth,
%``The Inflationary Universe: A Possible Solution to the Horizon and Flatness Problems,''
Phys. Rev. D \textbf{23}, 347-356 (1981)

\bibitem{Wang:2003yr}
B.~l.~Wang, H.~y.~Liu and L.~x.~Xu,
%``Accelerating universe in a big bounce model,''
Mod. Phys. Lett. A \textbf{19}, 449-456 (2004).

%\cite{Brandenberger:2012zb}
\bibitem{Brandenberger:2012zb}
R.~H.~Brandenberger,
%``The Matter Bounce Alternative to Inflationary Cosmology,''
[arXiv:1206.4196 [astro-ph.CO]].

%\cite{Novello:2008ra, deHaro:2012xj, Battefeld:2014uga, Ijjas:2016tpn}
\bibitem{Novello:2008ra}
M.~Novello and S.~E.~P.~Bergliaffa,
%``Bouncing Cosmologies,''
Phys. Rept. \textbf{463}, 127-213 (2008).
%doi:10.1016/j.physrep.2008.04.006
%[arXiv:0802.1634 [astro-ph]].

%\cite{deHaro:2012xj}
\bibitem{deHaro:2012xj}
J.~de Haro,
%``Does loop quantum cosmology replace the big rip singularity by a non-singular bounce?,''
JCAP \textbf{11}, 037 (2012).

%\cite{Battefeld:2014uga}
\bibitem{Battefeld:2014uga}
D.~Battefeld and P.~Peter,
%``A Critical Review of Classical Bouncing Cosmologies,''
Phys. Rept. \textbf{571}, 1-66 (2015).
%doi:10.1016/j.physrep.2014.12.004
%[arXiv:1406.2790 [astro-ph.CO]].

%\cite{Ijjas:2016tpn}
\bibitem{Ijjas:2016tpn}
A.~Ijjas and P.~J.~Steinhardt,
%``Classically stable nonsingular cosmological bounces,''
Phys. Rev. Lett. \textbf{117}, no.12, 121304 (2016).

\bibitem{Battye:2006mb}
R.~A.~Battye and A.~Moss,
%``Anisotropic perturbations due to dark energy,''
Phys. Rev. D \textbf{74}, 041301 (2006)
%doi:10.1103/PhysRevD.74.041301

\bibitem{Cai:2014bea}
Y.~F.~Cai,
%``Exploring Bouncing Cosmologies with Cosmological Surveys,''
Sci. China Phys. Mech. Astron. \textbf{57}, 1414-1430 (2014)
%doi:10.1007/s11433-014-5512-3

\bibitem{Khoury:2001wf}
J.~Khoury, B.~A.~Ovrut, P.~J.~Steinhardt and N.~Turok,
%``The Ekpyrotic universe: Colliding branes and the origin of the hot big bang,''
Phys. Rev. D \textbf{64}, 123522 (2001)
%doi:10.1103/PhysRevD.64.123522

\bibitem{Cai:2012va}
Y.~F.~Cai, D.~A.~Easson and R.~Brandenberger,
%``Towards a Nonsingular Bouncing Cosmology,''
JCAP \textbf{08}, 020 (2012).

\bibitem{Ashtekar:2006wn}
A.~Ashtekar, T.~Pawlowski and P.~Singh,
%``Quantum Nature of the Big Bang: Improved dynamics,''
Phys. Rev. D \textbf{74}, 084003 (2006)

\bibitem{Brandenberger:2016vhg}
R.~Brandenberger and P.~Peter,
%``Bouncing Cosmologies: Progress and Problems,''
Found. Phys. \textbf{47}, no.6, 797-850 (2017)
%doi:10.1007/s10701-016-0057-0

\bibitem{Cai:2007qw}
Y.~F.~Cai, T.~Qiu, Y.~S.~Piao, M.~Li and X.~Zhang,
%``Bouncing universe with quintom matter,''
JHEP \textbf{10}, 071 (2007).
%doi:10.1088/1126-6708/2007/10/071
%[arXiv:0704.1090 [gr-qc]].

%\cite{Cai:2008ed}
\bibitem{Cai:2008ed}
Y.~F.~Cai and X.~Zhang,
%``Evolution of Metric Perturbations in Quintom Bounce model,''
JCAP \textbf{06}, 003 (2009).
%doi:10.1088/1475-7516/2009/06/003
%[arXiv:0808.2551 [astro-ph]].

%\cite{Bamba:2008ut}
\bibitem{Bamba:2008ut}
K.~Bamba, S.~Nojiri and S.~D.~Odintsov,
%``The Universe future in modified gravity theories: Approaching the finite-time future singularity,''
JCAP \textbf{10}, 045 (2008)

%\cite{Odintsov:2014gea}
\bibitem{Odintsov:2014gea}
S.~D.~Odintsov and V.~K.~Oikonomou,
%``Matter Bounce Loop Quantum Cosmology from $F(R)$ Gravity,''
Phys. Rev. D \textbf{90}, no.12, 124083 (2014).
%doi:10.1103/PhysRevD.90.124083
%[arXiv:1410.8183 [gr-qc]].

\bibitem{Odintsov:2015zza}
S.~D.~Odintsov and V.~K.~Oikonomou,
%``Bouncing cosmology with future singularity from modified gravity,''
Phys. Rev. D \textbf{92}, no.2, 024016 (2015).

%\cite{Shaily:2024rjq, Singh:2018xjv, Singh:2023gxd}
\bibitem{Singh:2022gln}
J.~K.~Singh, Shaily and K.~Bamba,
%``Bouncing universe in modified Gauss\textendash{}Bonnet gravity,''
Chin. J. Phys. \textbf{84}, 371-380 (2023).

\bibitem{Shaily:2024rjq}
Shaily, J.~K.~Singh and A.~Singh,
%``Bouncing Cosmology in f(R,G) Gravity with Thermodynamic Analysis,''
Fortsch. Phys. \textbf{72}, no.6, 2300244 (2024).
%doi:10.1002/prop.202300244

%\cite{Singh:2018xjv}
\bibitem{Singh:2018xjv}
J.~K.~Singh, K.~Bamba, R.~Nagpal and S.~K.~J.~Pacif,
%``Bouncing cosmology in $f(R,T)$ gravity,''
Phys. Rev. D \textbf{97}, no.12, 123536 (2018).
%doi:10.1103/PhysRevD.97.123536
%[arXiv:1807.01157 [gr-qc]].

%\cite{Singh:2023gxd}
\bibitem{Singh:2023gxd}
J.~K.~Singh, Shaily, A.~Singh, A.~Beesham and H.~Shabani,
%``A non-singular bouncing cosmology in f(R,T) gravity,''
Annals Phys. \textbf{455}, 169382 (2023).
%doi:10.1016/j.aop.2023.169382
%[arXiv:2304.11578 [gr-qc]].

\bibitem{Singh:2022jue}
J.~K.~Singh, H.~Balhara, K.~Bamba and J.~Jena,
%``Bouncing cosmology in modified gravity with higher-order curvature terms,''
JHEP \textbf{03}, 191 (2023).
%doi:10.1007/JHEP03(2023)191
%[arXiv:2206.12423 [gr-qc]].

%\cite{Agrawal:2021rur}
\bibitem{Agrawal:2021rur}
A.~S.~Agrawal, L.~Pati, S.~K.~Tripathy and B.~Mishra,
%``Matter bounce scenario and the dynamical aspects in f(Q,T) gravity,''
Phys. Dark Univ. \textbf{33}, 100863 (2021).

%\cite{Lalke:2023cia}
\bibitem{Lalke:2023cia}
A.~R.~Lalke, G.~P.~Singh and A.~Singh,
%``Late-time acceleration from ekpyrotic bounce in f(Q,T) gravity,''
Int. J. Geom. Meth. Mod. Phys. \textbf{20}, no.08, 2350131 (2023).

\bibitem{Singh:2024tur}
A.~Singh, Shaily, J.~K.~Singh and E.~G{\"u}dekli,
%``Cosmic reverberations on a constrained f(Q,T)-model of the Universe,''
Annals Phys. \textbf{483}, 170274 (2025).
%%%%%%%%%

\bibitem{Paliathanasis:2016vsw}
A.~Paliathanasis, J.~D.~Barrow and P.~G.~L.~Leach,
%``Cosmological Solutions of $f(T)$ Gravity,''
Phys. Rev. D \textbf{94}, no.2, 023525 (2016).

\bibitem{Duchaniya:2022rqu}
L.~K.~Duchaniya, S.~V.~Lohakare, B.~Mishra and S.~K.~Tripathy,
%``Dynamical stability analysis of accelerating f(T) gravity models,''
Eur. Phys. J. C \textbf{82}, no.5, 448 (2022).

%%%%%%%%%%%%%%%%%%%%%%%%%%%%%%
\bibitem{delaCruz-Dombriz:2018nvt}
{\'A}.~de la Cruz-Dombriz, G.~Farrugia, J.~L.~Said and D.~S{\'a}ez-Chill{\'o}n G{\'o}mez,
%``Cosmological bouncing solutions in extended teleparallel gravity theories,''
Phys. Rev. D \textbf{97}, no.10, 104040 (2018).
%doi:10.1103/PhysRevD.97.104040
%[arXiv:1801.10085 [gr-qc]].

\bibitem{Bajardi:2020fxh}
F.~Bajardi, D.~Vernieri and S.~Capozziello,
%``Bouncing Cosmology in f(Q) Symmetric Teleparallel Gravity,''
Eur. Phys. J. Plus \textbf{135}, no.11, 912 (2020).

%\cite{Agrawal:2022vdg}
\bibitem{Agrawal:2022vdg}
A.~S.~Agrawal, B.~Mishra and P.~K.~Agrawal,
%``Matter bounce scenario in extended symmetric teleparallel gravity,''
Eur. Phys. J. C \textbf{83}, no.2, 113 (2023).

%\cite{Gadbail:2023loj}
\bibitem{Gadbail:2023loj}
G.~N.~Gadbail, A.~Kolhatkar, S.~Mandal and P.~K.~Sahoo,
%``Correction to Lagrangian for bouncing cosmologies in f(Q) gravity,''
Eur. Phys. J. C \textbf{83}, no.7, 595 (2023).

\bibitem{Sharif:2024bwy}
M.~Sharif, M.~Zeeshan Gul and N.~Fatima,
%``A comprehensive study of cosmic dynamics in f(Q) theory,''
Chin. J. Phys. \textbf{91}, 66-83 (2024).

\bibitem{Koussour:2024wtt}
M.~Koussour and N.~Myrzakulov,
%``Bouncing cosmologies and stability analysis in symmetric teleparallel f(Q) gravity,''
Eur. Phys. J. Plus \textbf{139}, no.9, 799 (2024).

%\cite{Zhadyranova:2024hbc}
\bibitem{Zhadyranova:2024hbc}
A.~Zhadyranova, M.~Koussour and S.~Bekkhozhayev,
%``The dynamics of matter bounce cosmology in Weyl-type f(Q,T) gravity,''
Chin. J. Phys. \textbf{89}, 1483-1492 (2024).

\bibitem{Basilakos:2025olm}
S.~Basilakos, A.~Paliathanasis and E.~N.~Saridakis,
%``Equivalence of f(Q) cosmology with quintom-like scenario: The phantom field as effective realization of the non-trivial connection,''
Phys. Lett. B \textbf{868}, 139658 (2025).
%%%%%%%%%%%%%%%%%%%%%%%%%%%%%%%%

%\cite{Will:2014kxa}
\bibitem{Will:2014kxa}
C.~M.~Will,
%``The Confrontation between General Relativity and Experiment,''
Living Rev. Rel.  \textbf{17}, 4 (2014).
%doi:10.12942/lrr-2014-4
%[arXiv:1403.7377 [gr-qc]].
%2677 citations counted in INSPIRE as of 04 Jul 2025

\bibitem{BeltranJimenez:2019tme}
J.~Beltr{\'a}n Jim{\'e}nez, L.~Heisenberg, T.~S.~Koivisto and S.~Pekar,
%``Cosmology in $f(Q)$ geometry,''
Phys. Rev. D \textbf{101}, no.10, 103507 (2020).

\bibitem{Blagojevic:2002du}
M.~Blagojevic,
%``Gravitation and gauge symmetries,''
doi:10.1201/9781420034264

\bibitem{Proca:1936fbw}
A.~Proca,
%``Sur la theorie ondulatoire des electrons positifs et negatifs,''
J. Phys. Radium \textbf{7}, 347-353 (1936).

\bibitem{Weinberg:1972kfs}
S.~Weinberg,
%``Gravitation and Cosmology: Principles and Applications of the General Theory of Relativity,''
John Wiley and Sons, 1972, ISBN 978-0-471-92567-5.

\bibitem{Scholz:2017pfo}
E.~Scholz,
%``The unexpected resurgence of Weyl geometry in late 20-th century physics,''
Einstein Stud. \textbf{14}, 261-360 (2018).

\bibitem{Berezin:2022odj}
V.~A.~Berezin and V.~I.~Dokuchaev,
%``Weyl cosmology,''
Int. J. Mod. Phys. A \textbf{37}, no.20n21, 2243005 (2022).
%doi:10.1142/S0217751X22430059
%[arXiv:2203.04257 [gr-qc]].

\bibitem{BeltranJimenez:2019acz}
J.~Beltr{\'a}n Jim{\'e}nez and A.~Delhom,
%``Ghosts in metric-affine higher order curvature gravity,''
Eur. Phys. J. C \textbf{79}, no.8, 656 (2019).

%\cite{Odintsov:2016tar}
\bibitem{Odintsov:2016tar}
S.~D.~Odintsov and V.~K.~Oikonomou,
%``Deformed Matter Bounce with Dark Energy Epoch,''
Phys. Rev. D \textbf{94}, no.6, 064022 (2016).
%doi:10.1103/PhysRevD.94.064022
%[arXiv:1606.03689 [gr-qc]].

%\cite{Saridakis:2018fth}
\bibitem{Saridakis:2018fth}
E.~N.~Saridakis, S.~Banerjee and R.~Myrzakulov,
%``Bounce and cyclic cosmology in new gravitational scalar-tensor theories,''
Phys. Rev. D \textbf{98}, no.6, 063513 (2018).
%doi:10.1103/PhysRevD.98.063513
%[arXiv:1807.00346 [gr-qc]].

%\cite{Lai:2025efh}
\bibitem{Lai:2025efh}
J.~Lai and C.~Li,
%``Probing the bounce energy scale in bouncing cosmologies with pulsar timing arrays,''
Eur. Phys. J. C \textbf{85} (2025) no.6, 714.
%doi:10.1140/epjc/s10052-025-14428-y
%[arXiv:2504.19251 [astro-ph.CO]].
%5 citations counted in INSPIRE as of 06 Dec 2025

%\cite{Agrawal:2022ppe}
\bibitem{Agrawal:2022ppe}
A.~S.~Agrawal, S.~K.~Tripathy, S.~Pal and B.~Mishra,
%``Role of extended gravity theory in matter bounce dynamics,''
Phys. Scripta \textbf{97}, no.2, 025002 (2022).
%doi:10.1088/1402-4896/ac49b2
%[arXiv:2201.03783 [gr-qc]].

%\cite{Raychaudhuri:1953yv}
\bibitem{Raychaudhuri:1953yv}
A.~Raychaudhuri,
%``Relativistic cosmology. 1.,''
Phys. Rev. \textbf{98}, 1123-1126 (1955).
%doi:10.1103/PhysRev.98.1123

%\cite{Kontou:2020bta}
\bibitem{Kontou:2020bta}
E.~A.~Kontou and K.~Sanders,
%``Energy conditions in general relativity and quantum field theory,''
Class. Quant. Grav. \textbf{37}, no.19, 193001 (2020).

%\cite{Tripathy:2020atm}
\bibitem{Tripathy:2020atm}
S.~K.~Tripathy, B.~Mishra, S.~Ray and R.~Sengupta,
%``Bouncing universe models in an extended gravity theory,''
Chin. J. Phys. \textbf{71}, 610-622 (2021).
%doi:10.1016/j.cjph.2021.03.026
%[arXiv:2002.03787 [gr-qc]].

%\cite{Chandrasekhar:1964zz}
\bibitem{Chandrasekhar:1964zz}
S.~Chandrasekhar,
%``The Dynamical Instability of Gaseous Masses Approaching the Schwarzschild Limit in General Relativity,''
Astrophys. J. \textbf{140}, 417-433 (1964).
%[erratum: Astrophys. J. \textbf{140} (1964), 1342].
%doi:10.1086/147938
%723 citations counted in INSPIRE as of 17 Dec 2024

%\cite{Banerjee:2024inf}
\bibitem{Banerjee:2024inf}
A.~Banerjee, A.~Pradhan, B.~Dayanandan and A.~Ali,
%``The role of pressure anisotropy on quark stars in gravity\textquoteright{}s rainbow,''
Eur. Phys. J. C \textbf{84}, no.7, 730 (2024).
%doi:10.1140/epjc/s10052-024-13120-x
%1 citations counted in INSPIRE as of 22 Dec 2024

%\cite{Shaily:2025kua}
\bibitem{Shaily:2025kua}
Shaily, J.~K.~Singh, D.~Sethi, R.~Rani and K.~Bamba,
%``Bouncing cosmology and the dynamical stability analysis in f(R,Lm)-gravity,''
Nucl. Phys. B \textbf{1013}, 116854 (2025).
%doi:10.1016/j.nuclphysb.2025.116854
%0 citations counted in INSPIRE as of 19 May 2025


\bibitem{Barvinsky:2013mea}
A.~O.~Barvinsky,
%``Dark matter as a ghost free conformal extension of Einstein theory,''
JCAP \textbf{01}, 014 (2014).
%doi:10.1088/1475-7516/2014/01/014

\bibitem{BeltranJimenez:2016wxw}
J.~Beltran Jimenez, L.~Heisenberg and T.~S.~Koivisto,
%``Cosmology for quadratic gravity in generalized Weyl geometry,''
JCAP \textbf{04}, 046 (2016).
%doi:10.1088/1475-7516/2016/04/046

\bibitem{Lin:2020phk}
Y.~C.~Lin, M.~P.~Hobson and A.~N.~Lasenby,
%``Ghost- and tachyon-free Weyl gauge theories: A systematic approach,''
Phys. Rev. D \textbf{104}, no.2, 024034 (2021).
%doi:10.1103/PhysRevD.104.024034

\bibitem{Gibbons:2003yj}
G.~W.~Gibbons,
%``Phantom matter and the cosmological constant,''
[arXiv:hep-th/0302199 [hep-th]].
%235 citations counted in INSPIRE as of 19 Aug 2022

\bibitem{Barrow:1988xh}
J.~D.~Barrow and S.~Cotsakis,
%``Inflation and the Conformal Structure of Higher Order Gravity Theories,''
Phys. Lett. B \textbf{214}, 515-518 (1988).
%doi:10.1016/0370-2693(88)90110-4

\bibitem{Shabani:2017lgy}
H.~Shabani and A.~H.~ziaie,
%``Late-time cosmological evolution of a general class of $f(\mathsf{R},\mathsf{T})$ gravity with minimal curvature-matter coupling,''
Eur. Phys. J. C \textbf{77}, no.8, 507 (2017).
%doi:10.1140/epjc/s10052-017-5077-1
%[arXiv:1703.06522 [gr-qc]].

\bibitem{Caldwell:1999ew}
R.~R.~Caldwell,
%``A Phantom menace?,''
Phys. Lett. B \textbf{545}, 23-29 (2002).
%doi:10.1016/S0370-2693(02)02589-3

%\cite{Cline:2003gs}
\bibitem{Cline:2003gs}
J.~M.~Cline, S.~Jeon and G.~D.~Moore,
%``The Phantom menaced: Constraints on low-energy effective ghosts,''
Phys. Rev. D \textbf{70}, 043543 (2004).
%doi:10.1103/PhysRevD.70.043543
%[arXiv:hep-ph/0311312 [hep-ph]].
%671 citations counted in INSPIRE as of 19 Aug 2022

%\cite{Nojiri:2003ag}
\bibitem{Nojiri:2003ag}
S.~Nojiri and S.~D.~Odintsov,
%``Effective equation of state and energy conditions in phantom/tachyon inflationary cosmology perturbed by quantum effects,''
Phys. Lett. B \textbf{571}, 1-10 (2003).
%doi:10.1016/j.physletb.2003.08.013
%[arXiv:hep-th/0306212 [hep-th]].
%224 citations counted in INSPIRE as of 19 Aug 2022

\end{thebibliography}
\end{document}